\documentclass[12pt]{iopart}
\usepackage{iopams,setstack,hvfloat,graphicx,amsfonts,amstext,latexsym}

\def \l{\left}
\def \r{\right}
\def \spc{\hspace{0.5em}}

\def \d{\text{d}}
\def \Max{{\hat {\rule{0em}{0.75em}\Box}\rule{0em}{0.75em}^\mu}_{\,\nu}}
\def \max{\hat {\Delta}^m_{\,\,\,n}}

\newcommand{\Path}[1]{#1}
\newcommand{\ignore}[1]{}

\newcommand{\tf}[2]{\frac{#1}{#2}}

\begin{document}

\title{Cosmological Perturbations in Flux Compactifications}
\author{Johannes U.\ Martin\footnote{Present address: University of
    Toronto, Toronto, ON, M5S 3H8, Canada.}}
\ead{jmartin@cita.utoronto.ca}

\begin{abstract}
  Kaluza--Klein compactifications with four-dimensional inflationary
  geometry combine the attractive idea of higher dimensional models
  with an attempt to incorporate four-dimensional early-time or
  late-time cosmology. We analyze the mass spectrum of cosmological
  perturbations around such compactifications, including the scalar,
  vector and tensor sector.  Whereas scalar perturbations were
  discussed before, the spectrum of vector and tensor perturbations is
  a new result of this paper.  Moreover, the complete analysis
  shows that possible instabilities of such compactifications are
  restricted to the scalar sector. The mass squares of the vector and
  tensor perturbations are all non-negative.  We discuss form fields
  with a non-trivial background flux in the extra space as matter
  degrees of freedom. They provide a source of scalar and vector
  perturbations in the effective four-dimensional theory.  We analyze
  the perturbations in Freund-Rubin compactifications. Although it can
  only be considered as a toy model, we expect the results to
  qualitatively generalize to similar configurations. We find that
  there are two possible channels of instabilities in the scalar
  sector of perturbations, whose stabilization has to be addressed in
  any cosmological model that incorporates extra dimensions and form
  fields. One of the instabilities is associated with the
  perturbations of the form field.
\end{abstract}
\maketitle

\section{Introduction}
\label{sec:intro}

In higher dimensional supergravity and string theories
compactifications of spacetimes are an essential tool to make
contact with the four-dimensional phenomenology. In the past the
majority of research was focused on the investigation of
compactifications to four-dimensional Minkowski or anti-de Sitter
spacetime. The Kaluza--Klein spectrum of such compactifications was
studied in detail to analyze the properties of the resulting
supergravity theories~\cite{Salam:1989fm}.

In contrary, cosmology motivates the study of effective
four-dimensional geometries that correspond to an expanding
universe. The successful cosmological model with quasi-de Sitter
epochs during inflation and dark energy domination today gradually
shifts the interest towards higher dimensional models that provide
possibilities for incorporating effective four-dimensional
cosmology. So far, realistic embeddings of cosmology/de Sitter
geometry into fundamental superstring/M-theories require tools
beyond simple low-energy supergravity
solutions~\cite{Maldacena:2000mw,Giddings:2001yu,Kachru:2003aw}.

Higher rank form fields are natural ingredients of higher
dimensional supergravity theories. In particular, in the attempt to
construct cosmological solutions, form fields play a crucial role.
They are utilized to stabilize the moduli fields associated with the
shape of the extra dimensions~\cite{Giddings:2001yu}. Furthermore,
non-trivial background fluxes in the extra dimensional space
generically induce exponential potentials for the volume modulus or
radion.  The applicability of such potentials for cosmological
quintessence and inflaton fields was recently investigated
in~\cite{Cornalba:2002fi,Frey:2002qc,Emparan:2003gg,Chen:2003dc,Neupane:2003pw}.
Commonly the dynamics are analyzed within the effective
four-dimensional theory, which has obvious limitations. Most
importantly, it requires a stable higher dimensional background
configuration\footnote{In alternative attempts to embed inflationary
  cosmology into higher dimensional theories, so-called s-brane
  solutions are used that are explicitly time-dependent, see e.g.\
  \cite{Ohta:2003ie}, which usually does not yield an effective
  four-dimensional description at all.}.  The stability of background
solutions is conveniently investigated by the study of linear
perturbations.

The analysis of perturbations in compactifications with effective
four-dimensional de Sitter geometry shows that it is not at all
generic for finding stable background configurations.  In
particular, realistic scenarios with a large number of extra
dimensions are plagued by two possible channels of instabilities in
the scalar sector of perturbations.

(i) The nature and dynamics of the tachyonic instability of the
lowest Kaluza--Klein state in the scalar spectrum---i.e., the volume
modulus---were investigated in~\cite{Frolov:2003yi,
Contaldi:2004hr}. The instability was recognized previously
in~\cite{Bousso:2002fi}. The generic contribution that arises from a
four-dimensional inflationary geometry with expansion rate $H$ is
given by
\begin{equation}
  \label{eq:tachyon}
  m^2_0=-\frac{12 q}{q+2} H^2\ ,
\end{equation}
where $q$ is the number of extra dimensions.  It is possible to
compensate this term with positive contributions from the curvature of
the internal space (if positive) and from stabilizing bulk fields,
e.g.\ scalar fields or form fluxes. If not stabilized this tachyon
indicates a non-linear reconfiguration of the
compactification~\cite{Martin:2003yh, Contaldi:2004hr}. Generic
late-time attractors are universal Kasner-like solutions, where the
internal dimensions shrink to zero size, or complete
decompactifications to a higher dimensional de Sitter space, where all
dimensions expand isotropically, or transitions towards regimes where
the de Sitter curvature is small and the tachyonic mode disappears.

(ii) It turns out that a second instability can arise from the
quadrupole and higher moments of Kaluza--Klein excitations $(l\geq
2)$ in the scalar sector.  Contrary to the instability of the volume
modulus that can be stabilized due to the interaction with matter
fields, this instability arises from the presence of non-trivial
background configurations of matter fields. Whereas the instability
of the volume modulus (that is, the Kaluza--Klein zero-mode)
preserves the spherical symmetry of the extra space, the second
instability indicates a deformation of the internal geometry.
Although the tachyonic mode was noticed in~\cite{Bousso:2002fi}, its
nature and consequences have not been explored so far and
non-perturbative examples for this instability have not been studied
yet. The range of values of the flux $c$ that allows for stable
compactifications shrinks with the number of extra dimensions. For
more than four extra dimensions, stable compactifications cannot be
found at all for this model. \ignore{Unlike
  the first instability, the second one cannot be removed by
  additional matter field degrees of freedom unless they have a
  coupling to the form flux as well.}

On the other hand, the mass spectrum of vector and tensor
perturbations does not reveal additional channels of instabilities.
For the vector sector, the lowest lying mass states are massless and
independent of the matter fields.  They transform in the adjoint
representation of the isometry group of the internal manifold and
their mass is protected by this local symmetry.  Unlike the scalar
sector of perturbations, the coupling between matter and metric
perturbations of the higher Kaluza--Klein vector modes does not lead
to additional instabilities in the sector of vector perturbations.

The lowest excitation of the tensor spectrum is the massless
four-dimensional graviton, whose mass is protected by
four-dimensional diffeomorphism invariance. Besides the massless
graviton a tower of positive Kaluza--Klein modes appears in the
effective theory, whose scaling is only sensitive to the properties
of the internal manifold.

Although the tachyonic instabilities render the discussion within
the effective four-dimensional theory invalid, they may trigger
interesting dynamics. The tachyonic scalars couple to the matter
fields in the effective four-dimensional theory. In Kaluza--Klein
compactifications the standard model fields of the effective
four-dimensional theory are considered to be zero-modes of the
higher dimensional theory. The unstable volume modulus couples
gravitationally to such fields, which might lead to interesting
phenomenological consequences such as tachyonic preheating of the
standard model fields. In simple braneworld models the preheating of
brane fields from the decaying radion was investigated
in~\cite{Kofman:2004tk}.

In this paper, we present the complete analysis of the mass spectrum
of perturbations around de Sitter vacua in a unified way. We
systematically treat scalar, vector and tensor perturbations. We
choose the model of Freund-Rubin compactifications, where all
spectra are obtained analytically.  Nevertheless, the qualitative
results are expected to be valid for general de Sitter
compactifications with flux stabilization that are frequently
discussed in the literature. In particular, we investigate the
resulting spectrum in the context of inflation. Light scalar modes
(with masses $m$ smaller than the inflation scale $H$) acquire a
scale invariant spectrum of perturbations after inflation and
therefore provide additional sources for cosmological perturbations.
The mass scale of Kaluza--Klein excitations of vector and tensor
modes is always larger than $H$, so that no contribution to
post-inflationary cosmology is expected.

The analysis follows closely the calculations done
in~\cite{Kim:1985ez,Bousso:2002fi} but goes substantially beyond
this for the case of vector perturbations and shape moduli fields,
which have not been addressed in the framework of de Sitter
compactifications before. The calculation of the complete spectrum
allows us to conclude that the tachyonic instability of de Sitter
compactifications is located in the scalar sector only.

The paper is structured as follows. In Sec.~\ref{sec:background}, we
set up the background solution for the de Sitter compactification.
The perturbations are introduced in Sec.~\ref{sec:perturbations}.
They are classified according to their transformation properties
with respect to the de Sitter isometry group. The equations of
motion for the perturbations are obtained and their spectra
calculated. In section~\ref{sec:applications}, we discuss possible
implications of the presence of light mass states for cosmological
models. Some technical details are shown
in~\ref{sec:Notation}--\ref{sec:residual}.

\section{The Background}
\label{sec:background}

We consider a $(p+q)$-dimensional product spacetime $M=M_p\times
M_q$. Like in Freund-Rubin compactification~\cite{Freund:1980xh}, we
allow for a $q$-form flux field that spontaneously compactifies $q$
of the spatial dimensions. For simplicity, we assume that the
compact space is a $q$-sphere---i.e., $M_q=S^q$ with radius $\rho$.
The generalization to any homogeneous space is straightforward. The
$p$-dimensional part of product space is a de Sitter space $dS_p$
with a curvature scale $H^2$. For cosmological applications $p=4$.
To compensate for the positive curvature of $M$, a cosmological
constant $\Lambda$ is introduced. We do not discuss the
microscopical origin of the cosmological constant, but we assume
that it is common to all dimensions. In principle, considering
quantum effects, one can expect additional contributions from the
Casimir energy to the cosmological constant of the compact space.
Its contribution however strongly depends on the choice and the
dimension of the compact manifold and we will assume that it can be
neglected in comparison to the overall cosmological constant.  The
theory we study is described by the action
\begin{equation}
  \label{eq:action}
   S=\int d^px d^qy
  \sqrt{|G|}\left\{\frac{1}{2}R-\frac{1}{2q!}F_{q}^2-\Lambda\right\}\ ,
\end{equation}
from which the general equations of motion for the metric and the $q$-form
field are derived
\begin{eqnarray}
  \fl R_{MN}=\frac{1}{(q-1)!}F_{MP_2\cdots P_q}{F_{N}}^{P_2\cdots
    P_q}\!-\!\frac{1}{q!}\frac{q-1}{p+q-2}F_q^2g_{MN}\!+\!\frac{2}{p+q-2}\Lambda g_{MN}\ ,\label{eq:eom1}\\
  \fl \nabla_{M}{F^{MP_2\cdots P_q}}=0\ .\label{eq:eom2}
\end{eqnarray}
The planar coordinate patch for $p$-dimensional de Sitter geometry is
parameterized by the coordinates $x^\mu$ and the $q$-dimensional
sphere with radius $\rho$ is labeled by the coordinates $y^m$
\begin{equation}
  \label{eq:line_back}
  ds^2=-dt^2+e^{2Ht}d\vec{x}^2+\rho^2 d\Omega_q^2\ .
\end{equation}
Throughout this work we use capital italic indices ${M,N,...}$ to
address all spacetime dimensions. Greek indices ${\mu,\nu,...}$ take
values in $0,...,p-1$ and correspondingly ${m,n,...}$ are used to
label the compactified dimensions. The static $q$-form flux that
supports this compactification is given by
\begin{equation}
\label{eq:flux_back}
  F_{P_1\cdots P_q}=\l\{\begin{array}{c c}
      c \epsilon_{p_1\cdots p_q} &\ ,\spc \text{if}\spc P_i=p_i\\
      \\
      0&\ ,\spc \text{otherwise,}
      \end{array}\r .
\end{equation}
where $\epsilon_{p_1\cdots p_q}$ is the completely antisymmetric
volume form of the $q$-sphere.

The background is characterized by four parameters: the de Sitter
scale $H$, the radius of compactification $\rho$, the flux strength
$c$ and the cosmological constant $\Lambda$. The equations of
motions~(\ref{eq:eom1}) and~(\ref{eq:eom2}) reduce to algebraic
constraints in this space of parameters
\begin{eqnarray}
    (q-1)\,\,\rho^{-2}-(p-1)\,\,H^2&=&c^2\ ,\nonumber\\
    (q-1)^2\rho^{-2}+(p-1)^2H^2&=&2\Lambda\ .\label{eq:param_back}
\end{eqnarray}
The last relation shows the requirement of a positive cosmological
constant that compensates for the curvatures of the spacetime. The
form field flux is an additional parameter that enriches the
dynamics twofold. Unlike the cosmological constant it is associated
with a field degree of freedom. It is natural to consider
perturbations of this field. Secondly, it allows one to create a
hierarchy between the Hubble scale $H$ and the size of the internal
manifold $\rho$, which is needed if one wants to apply this
background as approximate solution of the late-time universe, where
$H \ll \rho^{-1}$.

Dirac's quantization condition requires $c$ to be quantized. Although
we treat $c$ effectively as a continuous parameter, its discrete
nature is implied.

In the limit $c\rightarrow 0$, when the flux disappears, one
recovers the known results for standard Kaluza--Klein
compactifications. The range of values for the flux strength $c$ is
limited by the physical restriction of $H^2>0$.

From the first of the equations~(\ref{eq:param_back}) follows that
\begin{equation}
  \label{eq:c_constraint}
  c^2\rho^2\le (q-1)\ .
\end{equation}
The limit $c^2=c_{\text{max}}^2=(q-1)\rho^{-2}$ corresponds to
four-dimensional effective Minkowski space geometry with $H=0$.

\section{Perturbations around de Sitter compactifications}
\label{sec:perturbations}

\subsection{Classification of Perturbations and Gauge Fixing}
\label{sec:class_gauge}

In this section we analyze the dynamics of perturbations in the
background of Freund-Rubin compactifications introduced in the
previous section. First the metric perturbations are grouped into
scalar, vector and tensor perturbations. The gauge is fixed to
eliminate gauge degrees of freedom that correspond to the
infinitesimal coordinate transformations of the form
\begin{equation}
  \label{eq:inf_coord}
 x^M\rightarrow x^M+\xi^M(x)\ .
\end{equation}
Similarly perturbations for the matter fields are introduced, grouped
into scalars and vectors. Gauge degrees of freedom associated with the
transformation properties of the $(q-1)$-form gauge potential
$A_{q-1}$ are fixed.

The most general set of perturbations that respects the product
geometry of the chosen background spacetime is parameterized by the
following line element
\begin{eqnarray}
  \label{eq:line}
  \fl ds^2=\left[(1+2\hat\Psi)\gamma_{\mu\nu}+h_{(\mu\nu)}\right] dx^\mu
  dx^\nu+\l[(1+2\Phi)g_{mn}+h_{(mn)}\r]dy^mdy^n+V_{\mu n}dx^\mu dy^n\ ,
\end{eqnarray}
that contains the scalars $\hat \Psi$, $\Phi$ and $h_{(mn)}$, vectors
$V_{\mu n}$ and the $q$-dimensional traceless graviton $h_{(\mu\nu)}$.
Throughout this paper, indices in parentheses indicate the symmetrized
and traceless components of the tensorial object---i.e.,
$Q_{(mn)}=\frac{1}{2}\l(Q_{mn}+Q_{nm}\r)-\frac{1}{q}g_{mn}Q^m_m$.  It
is customary to further decompose the vectors and tensor into
longitudinal and transverse parts, e.g.\ $V_{\mu n}=V^{\perp}_{\mu
  n}+B_{n;\mu}$ and
$h_{(\mu\nu)}=h^{TT}_{\mu\nu}+F^{\perp}_{(\mu;\nu)}+2E_{;(\mu\nu)}$.
However, this decomposition is only relevant for the zero-modes---that
is, for the $y$-independent sector. The gauge choice that we impose
below in equations~(\ref{eq:dedonder}) has a residual gauge freedom that
allows us to impose $E=0$, $F^{\perp}_{\mu}=0$ and $B_n=0$ for the
$y$-independent perturbations, [cf.\
equations~(\ref{eq:res_constraint})]. For more details on the residual
gauge freedom we refer to the calculations in~\ref{sec:residual}.
In the massive---i.e., $y$-dependent ---sector these fields become the
longitudinal components of the vectors $V_{\mu n}$ and tensor
$h_{(\mu\nu)}$.

After compactification and integrating out the extra dimensions, the
graviton of the effective four-dimensional theory is not canonically
normalized unless the metric $\gamma_{\mu\nu}$ is rescaled by an
appropriate Weyl transformation, cf.~\ref{sec:weyl}. On the
linear level this amounts to the redefinition of the scalar
perturbation $\hat \Psi$ by the so-called Weyl shift
\begin{equation}
  \label{eq:weylshift}
  \Psi=\hat\Psi+\frac{q}{p-2}\Phi\ .
\end{equation}

To fix the gauge and to obtain equations of motion in a suitable form
the de Donder gauge conditions are imposed
\begin{equation}
  \label{eq:dedonder}
  {V^{\mu l}}_{|l}=0 \ ,\spc\spc\spc {h^{(m l)}}_{|l}=0 \ .
\end{equation}
The de Donder conditions are trivially satisfied for the
$y$-independent perturbations. Therefore, $V_{\mu n}$ consists of $q$
homogeneous vectors and $h_{(mn)}$ contains $q(q+1)/2-1$ homogeneous
scalars. For inhomogeneous perturbations, the de Donder gauge imposes
one condition on the vector fields $V_{\mu n}$ and $q$ conditions
among the scalars $h_{(mn)}$. Therefore, $V_{\mu n}$ represents $q-1$
massive vectors and $h_{(mn)}$ contains $q(q-1)/2 -1$ scalars.  The
missing vector and scalar modes are reshuffled to the longitudinal
polarizations of the graviton and the remaining vectors.  This has
been shown explicitly for compactifications on a $q$-dimensional torus
in~\cite{Han:1998sg}.

The gauge choice of equations~(\ref{eq:dedonder}) does not fix the gauge freedom
$\xi^M$ completely. The residual gauge freedom consists of functions
that satisfy the additional constraints
\begin{eqnarray}
  \label{eq:residual}
  \Delta \xi_\mu+{\xi^l}_{|l;\mu}&=&0\ ,\nonumber\\
  \xi_{(m|n)}&=&0 \nonumber \ .
\end{eqnarray}
We fixed the residual gauge that corresponds to the $y$-independent
coordinate transformations. The remaining residual gauge
transformations will be discussed in~\ref{sec:residual}.

Next we provide the notation for the form field perturbations and
the gauge choice analogous to the notation
of~\cite{Kim:1985ez,DeWolfe:2001nz}. The perturbations of the form
field strength are denoted with $f_{M_1 \cdots M_q}$. Unlike the
background, which only has non-vanishing components in the
directions of the compact space, the perturbations can fluctuate
freely in all directions. Locally, the perturbations of the $q$-form
field can be represented by a $(q-1)$-from potential---i.e.,
$f_{q}=\d a_{q-1}$.

Depending on the index structure of the potential $a_{q-1}$, it either
transforms as a scalar $a_{m_2\cdots m_q}$, vector $a_{\mu m_3\cdots
  m_q}$, or higher rank anti-symmetric tensor $a_{\mu \nu m_4 \cdots m_q}$.
In particular, interesting are scalars and vectors, since they mix
with the metric scalars and vectors. The decomposition of the field
strength into gauge field scalars and vectors is given by
\begin{eqnarray}
  \label{eq:potential}
  \fl f_{m_1\cdots m_q}=&q\, a_{[m_2\cdots m_q|m_1]}\ ,\nonumber\\
  \fl f_{\mu m_2\cdots m_q}=&q\,a_{[m_2\cdots
    m_q;\mu]}\,\,=&(-)^{q-1}(q-1)a_{\mu[m_2\cdots
    m_{q-1}|m_q]}+a_{m_2\cdots m_q;\mu}\ ,\\
  \fl f_{\mu\nu m_3\cdots m_q}=&q\, a_{[\nu m_3\cdots
    m_q;\mu]}=&a_{\nu m_3\cdots m_q;\mu}-a_{\mu m_3\cdots
    m_q;\nu}+(-)^{q-1}(q-2) a_{\mu\nu[m_3\cdots m_{q-1}|m_q]}\ .\nonumber
\end{eqnarray}
Like the electromagnetic gauge potential, the $(q-1)$-potential
$a_{q-1}$ consists of gauge degrees of freedom, represented by the
transformation $a_{q-1}\rightarrow a_{q-1}+\d\lambda_{q-2}$, where
$\lambda_{q-2}$ is an arbitrary $(q-2)$-form. To fix this gauge
freedom a Lorentz-like gauge condition is imposed
\begin{equation}
  \label{eq:lorentz}
  {a^{l}}_{m_3\cdots m_q|l}=0\ .
\end{equation}
A similar condition holds for the potentials with one or more
indices in the direction of the de Sitter space. On a compact
$q$-dimensional manifold without boundary any $n$-form $(n\le q)$
can be decomposed into an exact, a co-exact and a harmonic
$n$-form---i.e., $a_n=\d b_{n-1}+*\d* b_{n+1} + \beta^h_n$ where the
lower index indicates the rank of the form. The gauge choice in
equation~(\ref{eq:lorentz}) is equivalent to the statement that
$a_{q}$ is co-closed---that is, $*d* a_{q}=0$.  Therefore only the
co-exact and harmonic contribution survive in the general
decomposition.

The three-dimensional analogue of the above-described decomposition
is representing a vector $a^m$ whose divergence vanishes---i.e.,
${a^m}_{|m}=0$---as the curl of another vector,
$a^m=(\vec{\nabla}\!\times \vec{b}\,)^m =\epsilon^{mnp} b_{n|p}$.
Generalized to higher rank tensor objects the decomposition can be
written as
\begin{eqnarray}
  \label{eq:hodgedecomp}
  a_{m_2\cdots m_q}&=\sum\limits_I b^I(x) {\epsilon^m}_{m_2\cdots m_q}Y^I_{|m}(y)+\sum\limits_{h=1}^{b_1} \beta^h(x){\epsilon^{m}}_{m_2\cdots
    m_q} Y^h_m(y)\
  ,\nonumber\\
  a_{\mu m_3 \cdots m_q}&=\sum\limits_I b^I_\mu(x) {\epsilon^{mn}}_{m_3\cdots m_q}
  Y^I_{[n|m]}(y)+\sum\limits_{h=1}^{b_2} \beta^h_\mu(x){\epsilon^{mn}}_{m_3\cdots
    m_q} Y^h_{[mn]}(y)\ , \\
  &\spc\vdots\nonumber
\end{eqnarray}
where the first term in each line represents the generalization of the
curl and the second term summarizes the contribution from harmonic
forms on the compact space%
~\footnote{The number of independent
  harmonic $n$-forms $Y^h_{[m_1\cdots m_n]}(y)$ on a compact manifold
  depends on its $n$th Betti number---i.e., the number of non-trivial
  $n$-cycles.  On a $q$-sphere the only harmonic forms are the
  constant zero-forms and the volume form. The existence of additional
  scalars $\beta^h$ and vector fields $\beta^h_\mu$ requires harmonic
  one-forms $Y^h_m$ and two-forms $Y^h_{[mn]}$ as exist for
  example on a torus.}.
The $Y^I(y)$, $Y^I_m(y)$,$\cdots$, $Y^I_{[m_1\cdots m_q]}(y)$ are the
scalar, vector and tensor eigenforms of the Laplace-Beltrami operator
on the compact space---i.e., $(\d^{\dagger}\d+\d\d^{\dagger})
Y^I(y)=\kappa^I Y^I(y)$.

This completes the classification and gauge fixing of the
perturbations. In the following subsections the equations of motion
and spectra for the perturbations are derived. In the scalar sector,
the metric fluctuations $\Phi$ and $\Psi$ mix with the scalar from
the flux fluctuations $b$. Similarly, the fluctuations $V_{\mu n}$
couple to the flux perturbations $b_{\mu n}=\sum_I b^I_{\mu}(x)
Y^I_n(x)$ in the sector of vector fluctuations. The sector of tensor
modes is determined by the metric fluctuations $h_{(\mu\nu)}$ alone.

In what follows we ignore the harmonic sector, $\beta=0$, since it
gives rise to massless fields that are related to special topological
properties of the internal space. For the simple example of a
$q$-sphere, the only non-vanishing harmonic forms are the constant
scalar functions and the volume form, corresponding to the Betti
numbers $b_{0}=1$ and $b_q=1$. All other Betti numbers vanish.

\subsection{The Equations of Motion}
\label{sec:EOMS}

\subsubsection{The Linearized Einstein Equations.}
\label{sec:Einstein}

One set of dynamical equations for the perturbations are obtained
from the linearized version of the equations~(\ref{eq:eom1}). The
expression of the Ricci tensor is readily obtained from the general
formula for metric perturbations $\delta g_{MN}$:
\begin{eqnarray}
  \label{eq:dricci}
  \delta R_{MN}=&\frac{1}{2}\l[\rule{0em}{1.1em}\nabla_N\nabla_L
  \delta g^L_M+\nabla_M\nabla_L \delta g^L_N-\nabla_L\nabla^L \delta
  g_{MN}-\nabla_M\nabla_N \delta
  g^L_L\r .\nonumber\\&\l.+2 {R^L}_M \delta g_{NL}+2{R^L}_N \delta
  g_{ML}+2R^{K\,\,\,\,\,\,L}_{\,\,MN}\delta g_{KL}\r] ,
\end{eqnarray}
where $\delta
g_{\mu\nu}=(2\Psi-\frac{2q}{p-2}\Phi)\gamma_{\mu\nu}+h_{(\mu\nu)}$,
$\delta g_{\mu n}=V_{\mu n}$ and $\delta g_{mn}=2\Phi\,
g_{mn}+h_{(mn)}$, respectively.  The expression of the linearized
Ricci tensor is given in the set of
equations~(\ref{eq:ein1})--(\ref{eq:ein3}), where the gauge
conditions~(\ref{eq:dedonder}) were used for simplification.

The right-hand side of the field equations is expressed through a
linear combination $S_{MN}$ involving the energy--momentum tensor
$T_{MN}$:
\begin{equation}
  \label{eq:rhs}
 S_{MN}=T_{MN}-\frac{1}{p+q-2} T g_{MN}\ .
\end{equation}
The above expression~(\ref{eq:rhs}) is expanded up to first order in
the perturbations. The particular form~(\ref{eq:flux_back}) of the
background flux is used explicitly. This gives rise to various
contractions of the epsilon tensor $\epsilon_{p_1\cdots p_q}$ with
the perturbations of the form flux $f_{P_1\cdots P_q}$. These
contractions can be reduced to expressions in the perturbations $b$
and $b_{\mu n}$ only, that were introduced in the
equations~(\ref{eq:hodgedecomp}). Useful relations that simplify the
right-hand side of the linearized Einstein equations to the form
presented in the equations~(\ref{eq:ein4})--(\ref{eq:ein6}) are
collected in~\ref{sec:formequations}.
\begin{eqnarray}
 \fl  \delta
  R_{\mu\nu}=\frac{q}{p-2}(\Box+\Delta)\,\Phi\gamma_{\mu\nu}-(\Box+\Delta)\Psi\gamma_{\mu\nu}-(p-2)\Psi_{;\mu\nu}\nonumber\\+\frac{1}{2}\l[{h_{(\lambda\mu);\nu}}^\lambda+{h_{(\lambda\nu);\mu}}^\lambda-(\Box+\Delta) h_{(\mu\nu)}\r] \label{eq:ein1}\\
 \fl \delta R_{\mu
    n}=\l[\frac{p+q-2}{p-2}\Phi_{;\mu}-(p-1)\Psi_{;\mu}+\frac{1}{2} {h_{(\lambda\mu)}}^{;\lambda}\r]_{|n}\!+\frac{1}{2}\l[V^\lambda_{\,\,n;\mu\lambda}-(\Box+\Delta)V_{\mu n}+V^k_\mu\,R_{kn} \r] \nonumber\\\label{eq:ein2}\\
  \fl \delta R_{m
    n}=\l[\frac{2(p+q-2)}{p-2}\Phi-p\Psi\r]_{|(mn)}+\l[\frac{2(p+q-2)}{q(p-2)}\Delta\Phi-(\Box+\Delta)\Phi-\frac{p}{q}\Delta\Psi\r] g_{mn}\nonumber\\
  +\frac{1}{2}\l[V^\lambda_{m;\lambda|n}+V^\lambda_{n;\lambda|m}\r]\nonumber\\+\frac{1}{2}\l[-(\Box+\Delta)
  h_{(mn)}+R^l_m\, h_{(ln)}+R^l_n h_{(lm)}+2R^{k\spc\spc\spc l}_{\spc
    mn}h_{(kl)}\r]\ ,\label{eq:ein3}\\
 \fl \delta S_{\mu\nu}=\l[2q\l(\frac{q-1}{p+q-2}c^2\!-\!\frac{p\!-\!1}{p\!-\!2}H^2\r)\Phi\!+\!2(p\!-\!1)H^2\Psi\!-\!\frac{2(q-1)}{p+q-2}c\Delta b\r]\!\gamma_{\mu\nu}\nonumber\\+(p-1)H^2h_{(\mu\nu)}\label{eq:ein4}\\
  \fl \delta S_{\mu n}= c\l(b_{;\mu|n}+\Delta b_{\mu n}-
  b^l_{\mu|nl}\r)+(p-1)H^2V_{\mu n}
 \label{eq:ein5}\\
\fl \delta S_{m n}=\l\{2\frac{p-1}{p+q-2}\Delta
cb+2\l[(p-1)H^2-\frac{(q-1)(p-2)}{p+q-2}c^2\r]\Phi\r\}g_{mn}\nonumber\\+\l[c^2+(p-1)H^2\r]h_{(mn)}.\label{eq:ein6}
\end{eqnarray}

\subsubsection{The Linearized Maxwell Equations.}
\label{sec:Maxwell}

The linearized equations of motion of the form field are obtained
from the expansion of equation~(\ref{eq:eom2}) to first order in the
perturbations. The perturbed Christoffel symbols can be calculated
from the general formula
$\delta\Gamma^K_{MN}=\frac{1}{2}g^{KL}\l(\nabla_N \delta
g_{ML}+\nabla_M \delta g_{NL}-\nabla_L \delta g_{MN}\r)$. Again, the
covariant derivative is either constructed from $\gamma_{\mu\nu}$ or
$g_{mn}$ depending on the coordinate that is differentiated with
respect to. One finds
\begin{eqnarray}
   \fl 0=\l[{f_{\lambda p_2\cdots p_q}}^{;\lambda}+{f_{lp_2\cdots p_q}}^{|l}-c\delta
    \Gamma^l_{MN}g^{MN}\epsilon_{lp_2\cdots p_q}-(q-1) c g^{mn}\delta
    \Gamma^l_{p_2 n} \epsilon_{m l p_3\cdots
      p_q}\rule{0em}{2.5ex}\r]\epsilon^{p_1\cdots p_q}\ ,\label{eq:max1} \\
    \fl 0=\l({f_{\lambda\mu p_3\cdots p_q}}^{;\lambda}-{f_{\mu l
        p_3\cdots p_q}}^{|l}-c g^{mn}\delta\Gamma^l_{\mu n}\epsilon_{m
      l p_3\cdots p_q}\r)\epsilon^{p_1\cdots p_q}\ ,\label{eq:max2}\\
    \fl\spc\spc\vdots\nonumber\\
    \fl 0={f_{\lambda\mu_2\cdots\mu_q}}^{;\lambda}+{f_{l\mu_2\cdots
      \mu_q}}^{|l}\ .
\end{eqnarray}

It is convenient to factorize the $y$-dependence of the
perturbations in various tensor harmonics of the sphere. They
satisfy orthogonality relations and therefore break the system of
equations~(\ref{eq:ein1})--(\ref{eq:ein6}) into equations for each
representation $Y^I$.  Like the factorization in the
equations~(\ref{eq:hodgedecomp}), we introduce
\begin{eqnarray}
  \label{eq:tensordecomp}
  Q(x,y)&=&\sum_{I} Q^I(x) Y^I(y)\ ,\nonumber\\
  V_{\mu n}(x,y)&=&\sum_{I} V_\mu^I(x) Y^I_n(y)\ ,\nonumber\\
  h_{(mn)}(x,y)&=&\sum_{I} h^I(x) Y^I_{(mn)}(y)\ ,
\end{eqnarray}
where $Q$ collectively stands for the perturbations
$\{\Phi,\Psi,h_{(\mu\nu)}\}$ and $I$ is a collective index over the
eigenvalues of the corresponding tensor representation. We note
again that a pair of indices in parentheses indicates the
symmetrized and traceless components of the corresponding
second-rank tensor---i.e., $Y_{(mn)}g^{mn}=0$. Additionally, the
condition $Y^{(mn)}_{\spc\spc\spc\spc|m}=0$ holds, which is
compatible with the de Donder gauge condition~(\ref{eq:dedonder}).
In the following, we suppress the summation symbol and the index
$I$.

\subsection{Scalar Perturbations}
\label{sec:scalars}

First we calculate the mass spectrum for the scalar metric
perturbations $\Phi$ and $\Psi$ that couple to the scalar form field
perturbation $b$. To
find their equations of motion, we compare the coefficients in front
of the spherical harmonics $Y^I(y)$ in the Einstein
equations~(\ref{eq:ein1})--(\ref{eq:ein6}) and the Maxwell
equations~(\ref{eq:max1})--(\ref{eq:max2}). The first of these
equations is further simplified by taking the trace over $\mu$ and
$\nu$. One obtains the following set of equations:
\begin{eqnarray}
 \fl   \!\!\l\{\tf{1}{p}
  h_{(\kappa\lambda)}^{\spc\spc\spc;\kappa\lambda}+\frac{q}{p-2}\l(\Box+\Delta\r)\Phi+2q\l[\tf{p-1}{p-2}H^2-\tf{q-1}{p+q-2}c^2\r]\Phi+2\tf{q-1}{p+q-2}c\Delta b\r.&\nonumber\\
\l.-\l[2\tf{p-1}{p}\Box+\Delta+2(p-1)H^2\r]\Psi\r\}&\hspace{-3.5em}Y\,\,\spc\spc\,\,=0\label{eq:scalar1}\\
  \l\{\tf{1}{2}
  h_{(\lambda\mu)}^{\spc\spc\spc;\lambda}+\tf{p+q-2}{p-2}\Phi_{;\mu}-(p-1)\Psi_{;\mu}-cb_{;\mu}\r\}&\hspace{-3.5em}Y_{|n}\,\,\spc\,\,=0\label{eq:scalar2}\\
  \l\{2\frac{p+q-2}{p-2}\Phi-p\Psi\r\}&\hspace{-3.5em}Y_{|(mn)}=0\label{eq:scalar3}\\
\fl   \l\{ \l(\Box+\Delta\r)\Phi
+2\l[(p-1)H^2-\tf{(p-2)(q-1)}{p+q-2}c^2\r]\Phi+2\tf{p-1}{p+q-2}c\Delta
b\r.&\nonumber\\
\l.+\frac{1}{q}\l[p\Delta\Psi-\frac{2(p+q-2)}{p-2}\Delta\Phi\r]\r\}&\hspace{-3.5em}Y\,\spc\spc\,\,\,=0\label{eq:scalar4}\\
 \l\{c^2\l[p\Psi-2\tf{q(p-1)}{p-2}\Phi\r]+c(\Box+\Delta)b\r\}
&\hspace{-3.5em}Y_{|n}\,\,\spc\,\,=0\,.\label{eq:scalar5}
\end{eqnarray}
In what follows, the eigenvalues of the Laplace operator $\Delta$ are
denoted by $\lambda$. For a $q$-sphere $\lambda$ takes the values
$-l(l+q-1)$ for $l=0,1,2,...$ [i.e., $\rho^2\Delta Y=-l(l+q-1) Y$].

The set of equations~(\ref{eq:scalar1})--(\ref{eq:scalar5}) has to
be analyzed for three different cases.  (i) The general case, where
$Y_{|m}\neq 0$ and $Y_{|(mn)}\neq 0$, in which all five
equations~(\ref{eq:scalar1})--(\ref{eq:scalar5}) have to be
satisfied. (ii) The second case is a peculiarity of highly symmetric
compact manifolds, where $Y_{|m}\neq 0$ but $Y_{|(mn)}\equiv
Y_{|mn}-\frac{1}{q} \Delta Y g_{mn}=0$. For the $q$-sphere, the
spherical harmonics with angular momentum $l=1$ (i.e., the first
massive Kaluza--Klein excitations) satisfy this property. (iii) The
last case deals with the homogeneous perturbations, for which both
$Y_{|m}=0$ and $Y_{|(mn)}=0$.

\subsubsection{The General Case $(l>1$, $\lambda>q)$: $Y_{|m}\neq 0$ and $Y_{|(mn)}\neq 0$.}
Let us first consider the general ($y$-dependent) case.
Equation~(\ref{eq:scalar3}) can be used to eliminate $\Psi$.
Additionally, $h^{(\kappa\lambda)}_{\spc\spc\spc;\kappa\lambda}$ in
equation~(\ref{eq:scalar1}) is eliminated with
equation~(\ref{eq:scalar2}).  Then it is straightforward to show
that the first equation is a linear combination of
equations~(\ref{eq:scalar4}) and~(\ref{eq:scalar5}), which are the
dynamical equations for the perturbations $b$ and $\Phi$. We define
the following dimensionless quantities to shorten the notation:
\begin{eqnarray}
  \label{eq:def}
  A&=(q-1) c^2\rho^2\ ,\\
  B&=\l[\tf{p-2}{p+q-2}A-(p-1)(H\rho)^2 \r]=\l[\tf{q(p-1)}{p+q-2}c^2\rho^2\!-\!(q-1)\r],\nonumber
\end{eqnarray}
and denote the eigenvalues of the Laplacian on the $q$-sphere with
$\lambda$, $\Delta Y^l = -\lambda^l\rho^{-2} Y^l$, with
$\lambda^l=l(l+q-1)$.  Finally one obtains the dynamical system
\begin{eqnarray}
  \rho^2\Box\l(
  \begin{array}{c}
    c b\\
    \Phi
  \end{array}
\r)=
\l(
\begin{array}{c c}
  \lambda & 2\,A\\
  2\tf{p-1}{p+q-2}\lambda & \lambda+2B
\end{array}
\r)
\l(
  \begin{array}{c}
    c b\\
    \Phi
  \end{array}
\r)\ .\label{eq:scalar_eom}
\end{eqnarray}
Next the system~(\ref{eq:scalar_eom}) is diagonalized. The mass
spectrum and mass eigenstates are given by
\begin{eqnarray}
  \label{eq:scalar_mass}
  \rho^2m^2_\pm&=\lambda+B\pm\sqrt{B^2+\tf{4(p-1)}{p+q-2}A\lambda}\
  ,\\
   X_+&\propto-2\tf{p-1}{p+q-2}\lambda
   cb-\l[\sqrt{B^2+\tf{4q(p-1)}{p+q-2}A\lambda}+B\r] \Phi\ ,\nonumber\\
   X_-&\propto\,\,\,\,2\tf{p-1}{p+q-2}\lambda cb-\l[\sqrt{B^2+\tf{4q(p-1)}{p+q-2}A\lambda}-B\r] \Phi\ .\nonumber
\end{eqnarray}

\subsubsection{The Case $(l=1$, $\lambda=q)$: $Y_{|m}\neq0$ and
  $Y_{|(mn)}=0$.}

For this case, the algebraic relation~(\ref{eq:scalar3}) between
$\Phi$ and $\Psi$ cannot be imposed anymore. Again,
$h_{(\kappa\lambda)}^{\spc\spc\spc;\kappa\lambda}$ is eliminated from
the equations~(\ref{eq:scalar1}) and~(\ref{eq:scalar2}). $\Psi$ is
determined from the last equation~(\ref{eq:scalar5}). The remaining
two equations lead to identical equations of motion for the variable
$\chi=\Phi+\frac{1}{c^2\rho^2} cb$
\begin{equation}
  \label{eq:lambdagleichq}
  \rho^2\Box \chi=\l(2\frac{q(p-1)}{p+q-2}c^2\rho^2+q\r)\chi\ .
\end{equation}
If the general spectrum~(\ref{eq:scalar_mass}) is specialized to
the case $\lambda=q$, one finds that $X_+\propto \chi$ and
$m^2_+|_{\lambda=q}=m^2_\chi$. As argued in~\cite{Kim:1985ez}, the
second mode $X_-\propto cb$ that corresponds to the negative branch
of~(\ref{eq:scalar_mass}) is gauged away by the enlarged residual
gauge symmetry for the case $\lambda=q$ ($l=1$), cf.~\ref{sec:residual}.

\subsubsection{The Homogeneous Case $(l=0$, $\lambda=0)$: $Y_{|m}=0$ and
  $Y_{|(mn)}=0$.}

In the homogeneous case, the residual gauge freedom is enhanced by
the $p$-dimensional diffeomorphism invariance, $ x^\mu \rightarrow
x^\mu+\xi^\mu(x)$. This allows us to impose the transverse condition
on $h_{(\mu\nu)}$---i.e.,
$h^{(\mu\nu)}_{\spc\spc\spc;\mu}=0$---which is not valid for
inhomogeneous modes as will be shown in Sec.~\ref{sec:tensors}. This
difference in the properties of the homogeneous and inhomogeneous
graviton can be understood as follows: the homogeneous (massless)
graviton has $p(p-3)/2$ polarizations and its transverse and
traceless polarizations completely decouple from the scalar and
vector sector of perturbations. The inhomogeneous (massive) graviton
acquires $p-1$ longitudinal polarizations from the massive scalar
and vector sector. Therefore, the physical massive graviton is a
linear combination of $h_{(\mu\nu)}$ and the scalar and vector
modes. The explicit nature of this combination depends on the choice
of gauge.  For our gauge the physical massive graviton is given by
$\phi_{(\mu\nu)}$ that is introduced in
equation~(\ref{eq:phys_graviton}).

Additionally, for the case $\lambda=0$, the scalar Maxwell
equation~(\ref{eq:scalar5}) is a trivial identity. Therefore, the
homogeneous component of $b$ is unphysical and can be set to zero.
The absence of a zero-mode perturbation of the form flux respects
the quantized nature of $c$. Higher multipole perturbations average
to zero when integrated over the entire sphere; therefore the total
flux $c$ remains unchanged. Similarly, the
equations~(\ref{eq:scalar2}) and~(\ref{eq:scalar3}) are trivial
identities. One additional constraint arises from the off-diagonal
$(\mu\nu)$ Einstein equations. The massless graviton only decouples
from the scalar sector if $\Psi=0$. In the notation of the
paper~\cite{Contaldi:2004hr}, this constraint translates into
$\Psi_\text{\sc
  ckp}=-\tf{q}{2}\Phi_\text{\sc ckp}$, where the index {\sc ckp}
denotes the variables for scalar perturbations used
in~\cite{Contaldi:2004hr}.

The remaining equation for the variable $\Phi$ reduces to
\begin{equation}
  \label{eq:zeromode}
  \rho^2\Box \Phi=2\l[\tf{q(p-1)}{p+q+2}c^2\rho^2-(q-1)\r]\Phi\ .
\end{equation}
This corresponds to the positive branch in the general
formula~(\ref{eq:scalar_mass}) if $B>0$ and to the negative branch
in the case when $B<0$, giving a mass for the homogeneous mode
$\Phi$:
\begin{eqnarray}
  \label{eq:radion}
  \rho^2 m^2_0&=-2(q-1)+\frac{2q(p-1)}{p+q-2}c^2\rho^2\nonumber\\
&=-2(p-1)H^2\rho^2+\frac{2(p-2)(q-1)}{p+q-2}c^2\rho^2\ .
\end{eqnarray}
In the limit where the flux is turned off, we reproduce the result
found in~\cite{Contaldi:2004hr} of a tachyonic mode $m^2_0=-6H^2$ for
$p=4$. Remarkably, the mass is independent of the number of internal
dimensions.

\subsubsection{Instabilities from the Scalar Sector $(p=4)$.}

The appearance of a tachyonic mode in the spectrum of perturbations
signals the instability of the background configuration. In what
follows, we collect all possible channels of instabilities. It turns
out that the tachyonic modes all reside in the
scalar sector of the perturbations. In equation~(\ref{eq:radion}), we
have seen already the tachyonic nature of the homogeneous mode $\Phi$
that is generic if the flux is below a threshold value
$c^2_\text{th}=\frac{(p-1)(p+q-2)}{(q-2)(p-2)}H^2$ or in other words if
the inflation scale $H$ is higher than the flux $c$ that stabilizes
the compactification.

However, as the authors of~\cite{Bousso:2002fi} showed, the
homogeneous mode is not the only source for instabilities. If the
negative branch of the general formula~(\ref{eq:scalar_mass}) is
analyzed for the higher Kaluza--Klein excitations---i.e., $l\geq
2$---one finds additional tachyons above a flux critical
$c^2_\text{cr}$. In Tab.~\ref{tab:compare} the threshold flux for
the volume modulus, $l=0$ and the critical flux for the negative
branch of $l=2$ are calculated for various numbers $q$ of extra
dimensions and $p=4$. Stable compactifications only exist for fluxes
higher than the threshold flux $c_\text{th}$ and lower than the
critical flux $c^2_\text{cr}$. Consequently, stable configurations
are found for $q=2$ and $q=3$ extra dimensions for all fluxes larger
than $c_\text{th}$. In $q=4$ extra dimensions, stable
compactifications are only achieved for a range of fluxes
$c_\text{th}<c<c_\text{cr}$ and no stable compactifications are
possible for more than four extra dimensions.

\begin{table}[h]
  \centering
  \fbox{%
  \begin{tabular}[t]{c | c c}
    & $\frac{c^2_\text{th}}{c^2_\text{max}}$
    &$\frac{c^2_\text{cr}}{c^2_\text{max}}$ $(l=2)$\vspace{0.5ex}\\
\hline
$q=2$ &$\frac{2}{3}=0.\bar 6$& $\infty$\\
$q=3$ &$\frac{5}{9}=0.\bar 5$& $\frac{5}{3}=1.\bar
6$\\
$q=4$ &$\frac{1}{2}=0.5$& $\frac{2}{3}=0.\bar 6$\\
$q=5$ &$\frac{7}{15}=0.4\bar 6$& $\frac{7}{18}=0.3\bar
8$\\
$q=6$ &$\frac{4}{9}=0.\bar 4$& $\frac{4}{15}=0.2\bar 6$\\
$q=7$ &$\frac{3}{7}\approx0.43$ & $\frac{1}{5}=0.2$
  \end{tabular}
  }
  \caption{The threshold flux $c_\text{th}$ and the critical flux
    $c_\text{cr}$ for $p=4$ and various numbers of extra dimensions $q$. Stable compactifications are only admitted if $c_\text{th}<c<c_\text{cr}$. Therefore, only for $q=2,3$ and $4$ stable configurations are possible.}
  \label{tab:compare}
\end{table}

The mass spectrum of the scalar perturbations $\Phi$ and $b$ is
summarized in Fig.~\ref{fig:scalar_spectrum} for the number of extra
dimensions $q=3,4,$ and $5$. The mass of the zero-mode (denoted by
$0^\text{th}$ KK) is given by equation~(\ref{eq:radion}), the mass
for the special mode $l=1$ (denoted by $1^\text{st}$ KK) is obtained
from equation~(\ref{eq:lambdagleichq}) and all the other excitations
are graphical representations of the two branches $m_\pm$ in
equation~(\ref{eq:scalar_mass}). The threshold flux $c_\text{th}$
and the critical flux $c_{cr}$ are indicated by the vertical lines.

\subsubsection{Shape Moduli Dynamics.}
The dependence on $y$ of the metric perturbations $h_{(mn)}$ is
factorized according to equation~(\ref{eq:tensordecomp}).  The
dynamics of the shape moduli is obtained from the coefficient in
front of $Y_{(mn)}$ in the $(mn)$ Einstein equations. Using the
background equations~(\ref{eq:param_back}), the equation simplifies
to
\begin{equation}
  \label{eq:shape}
  \rho \Box h^I=\lambda^I h^I\ ,
\end{equation}
where $\lambda^I$ are the eigenvalues of the Laplace operator on
symmetric and traceless tensors on $S^q$---that is, $\rho^2 \Delta
Y^I_{(mn)}=-\lambda^I Y^I_{(mn)}$ with $\lambda^I=l(l+q-1)-2$ and
$l\ge 2$. Obviously, the sphere is stable against shape moduli
perturbations. The shape moduli fields have a positive Kaluza--Klein
spectrum, which only depends on the geometry of the internal space.
It is not affected by the details of the de Sitter geometry.
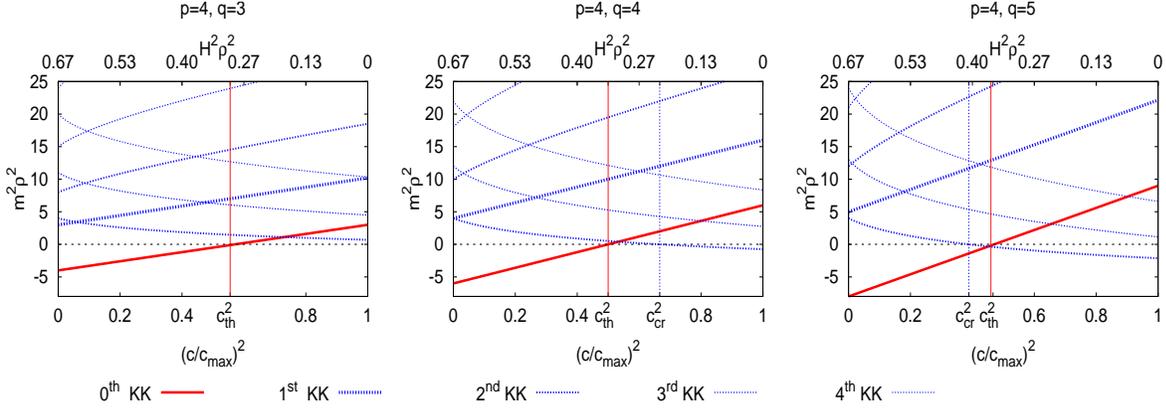
\hvFloat[nonFloat=false,capWidth=w]{figure}{\includegraphics[width=\textwidth,height=30ex,viewport=5 5 1080 325,clip]{\Path{scalar_spectrum_complete}}}{The
  mass spectrum for the scalar perturbations $\Phi$ and $b$ as a
  function of the $q$-form flux $c$ for various numbers of extra
  dimensions $q=3,4$ and $5$. For fluxes below the value
  $c_\text{th}$ the homogeneous volume modulus (solid, red line) has
  a tachyonic mass and renders the compactification unstable. For
  fluxes above $c_\text{cr}$ the negative branch of the second
  Kaluza--Klein excitation becomes tachyonic, again indicating an
  unstable background configuration.}{fig:scalar_spectrum}

\subsubsection{Effective 4D Theory.}

It is instructive to analyze the dynamics of the homogeneous scalar
mode $\Phi$ from the point of view of an effective four-dimensional
theory beyond the linear approximation. The analysis of
perturbations suggests the following ansatz for the scalar mode
$\Phi$:
\begin{equation}
  \label{eq:eff1}
  ds^2=e^{-\frac{2q}{p-2}\Phi(x)}\gamma_{\mu\nu}(x)dx^\mu
  dx^\nu+e^{2\Phi(x)}g_{mn}(y)dy^m dy^n\ .
\end{equation}
This particular choice of parameterization is compatible with the
results for the perturbations, i.e.\ $e^{2\Phi}=1+2\Phi+{\cal
  O}(\Phi^2)$, and it ensures that the effective four-dimensional
theory exhibits canonical Einstein gravity. The condition of flux
quantization requires that $c$ is no longer a constant. It is also
promoted to a four-dimensional field
\begin{equation}
  \label{eq:eff2}
 c=c_0 e^{-q\Phi(x)}\ .
\end{equation}
The spacetime-dependent factor takes into account the change in the
volume of the internal space such that the total flux remains
constant. The background Maxwell equations (\ref{eq:eom2}) are still
trivially satisfied. With the equations (\ref{eq:eff1}) and
(\ref{eq:eff2}) the action of the system (\ref{eq:action}) takes the
following form
\begin{eqnarray}
\label{eq:eff3}
S=\frac{1}{2}\int d^px
d^qy\sqrt{|\gamma|g}\l\{\rule{0em}{2.5ex}R[\gamma]\r.\!&\!-\frac{q(p+q-2)}{p-2}(\nabla\Phi)^2\\
&\l.+R[g]e^{-2\frac{p+q-2}{p-2}\Phi}\!-\!2\Lambda
e^{-2\frac{q}{p-q}\Phi}\!-\!c_0^2e^{-2q\frac{p-1}{p-2}\Phi}\r\}\ ,\nonumber
\end{eqnarray}
where we neglected a boundary term proportional to $\Box \Phi$.
$R[\gamma]$ is the four-dimensional Ricci scalar and
$R[g]=q(q-1)\rho^{-2}$ is the Ricci scalar constructed with the
metric $g_{mn}(y)$. After integrating out the extra dimensions and
rescaling $\Phi\rightarrow \sqrt{\frac{q(p+q-2)}{p-2}}\Phi$ we
obtain the effective potential $V_\text{eff}$ for the canonically
normalized field $\Phi$:
\begin{eqnarray}
  \label{eq:eff4}
  V_\text{eff}(\Phi)=&2\Lambda
  e^{-2\sqrt{\frac{q}{(p-2)(p+q-2)}}\Phi}+c_0^2e^{-2(p-1)\sqrt{\frac{q}{(p-2)(p+q-2)}}\Phi}\nonumber\\
  &-q(q-1)\rho^{-2}e^{-2\sqrt{\frac{p+q-2}{q(p-2)}}\Phi}\ .
\end{eqnarray}
The stability of the background solution is determined by the shape of
this potential at $\Phi=0$.  The parameters $\Lambda$, $c_0$, $H$ and
$\rho$ are related by the background equations of motion
(\ref{eq:param_back}). If Eqs.~(\ref{eq:param_back}) are taken into
account then the potential has the following properties at $\Phi=0$
\begin{eqnarray}
  \label{eq:eff5}
  \hspace{0.8em}V'_\text{eff}&=0\nonumber\\
  \frac{1}{2}V''_\text{eff}&=-2(p-1)H^2+2\frac{(p-2)(q-1)}{p+q-2}c_0^2=2m^2_0\ ,
\end{eqnarray}
which is equivalent to the results obtained in Eq.~(\ref{eq:radion}).

\begin{figure}[h]
  \centering
 \includegraphics[width=0.5\textwidth]{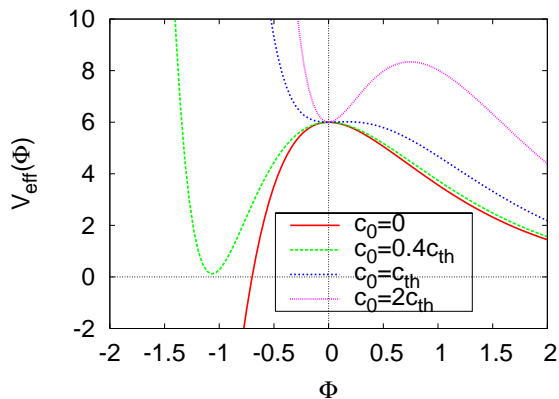}
 \caption{The
   effective potential $V_\text{eff}$ $(p=4$, $q=3)$ of the mode
   $\Phi$ for various values of fluxes. In the absence of fluxes
   $c_0=0$, the background solution is unstable. For fluxes
   $0<c_0<c_\text{th}$ a minimum is formed, to which the system can
   evolve, and for $c_0>c_\text{th}$ the background solution is
   stable.}
\label{fig:eff}
\end{figure}

However, the effective potential (\ref{eq:eff4}) allows us to analyze
the dynamics beyond the stability of the background solution. In
Fig.~\ref{fig:eff} we illustrate the effective potential for various
values of the flux $c_0$. It is clearly visible that in the case of
vanishing flux, i.e.\ $c_0=0$, the background solution corresponds to an
unstable maximum. It can either evolve towards $\Phi\rightarrow
-\infty$, which corresponds to the collapse of the internal manifold,
or towards $\Phi\rightarrow \infty$, which corresponds to the
decompactification of the internal space, cf.\ \cite{Contaldi:2004hr}.
For flux values $0<c_0<c_\text{th}$ a second minimum occurs at a
finite but smaller size of the internal space. As analyzed
in~\cite{Krishnan:2005su}, this corresponds to transitions of the
system towards a stable anti-de Sitter or de Sitter configuration. For
flux values of $c_0>c_\text{th}$ a minimum at $\Phi=0$ is formed and the
background solution is classically stable.

\subsection{Vector Perturbations}
\label{sec:vectors}

The mass spectrum for vector perturbations of the effective
$p$-dimensional theory is obtained from the equations of motion for
the vector perturbations $b_{\mu n}$ and $V_{\mu n}$. They appear as
coefficients in front of the vector harmonics $Y^I_n(y)$.  The system
of vector perturbations needs to be diagonalized to find the spectrum
and the mass eigenstates. They are linear combinations of the metric
vector perturbations $V_{\mu n}$ and the form flux vector
perturbations $b_{\mu n}$.

We list the equations of motion for the vector modes obtained from the
Einstein and Maxwell equations~(\ref{eq:ein1})--(\ref{eq:max2}).
The form field is expressed in terms of its gauge
potential~(\ref{eq:potential}). The gauge potential is further
decomposed according to the equations~(\ref{eq:hodgedecomp}) and the
relations~(\ref{eq:formulae}) are used for simplification
\begin{eqnarray}
  \l\{\l[
  \Max V_{\mu}+2(p-1)H^2V_{\nu}\r]\delta^m_n+(V_{\mu}+2cb_\nu)\max\r\}Y_{m}&=0\ ,\label{eq:vector1}\\
  \l\{c^2 {V^{\lambda}}_{;\lambda}\delta^m_n-\max c{b^{\lambda}}_{;\lambda}\r\}Y_{m}&=0\
  ,\label{eq:vector2}\\
  \l\{\l[-c\Max b_{\mu}+c^2 V_{\nu}\r]\delta^m_n-c\max b_{\nu}\r\}Y_m&=0\ ,\label{eq:vector3}
\end{eqnarray}
where $\Max=\Box\delta^{\,\mu}_{\,\,\nu}-\nabla^\mu\nabla_\nu$
represents the ordinary Maxwell operator and similarly $\max=\Delta
\delta^m_{\,\,n}-\nabla^m\nabla_n$, which up to a sign equals to the
action of the Laplace--Beltrami operator $(\d^\dagger
\d+\d\d^\dagger)$ on the one-forms $Y_m$.

The second equation~(\ref{eq:vector2}) follows from the third
one~(\ref{eq:vector3}) by differentiation. The two dynamical
equations~(\ref{eq:vector1}) and~(\ref{eq:vector3}) for the
physical vector modes $V_{\mu n}$ and $b_{\mu n}$ can be recast
compactly
\begin{eqnarray}
 \Max \!\l(\!
  \begin{array}{c}
    V^I_{\mu}\\
    c b^I_{\mu}
  \end{array}\!\r)\!Y^I_n\!=\!\l(\!
  \begin{array}[c]{c c}
    \kappa^I-2(p-1)H^2 & 2 \kappa^I\rho^{-2}\\
    c^2 & \kappa^I\rho^{-2}
  \end{array}\!\r)\!\!\l(\!
  \begin{array}[c]{c}
    V^I_\nu\\
    c b^I_\nu
  \end{array}\!\r)\!Y^I_n ,
\end{eqnarray}
where $\kappa^I$ are the eigenvalues of the Laplace--Beltrami
operator, $\rho^2\max Y_m^I(y)=-\kappa^I Y_n^I(y)$, that take values
in $(k+1)(q+k-2)$ for $k\ge 1$. The eigenvalues of the mass matrix
are given by
\begin{eqnarray}
  \fl \rho^2 m^2_\pm&=
  \kappa-(q-1)+c^2\rho^2\pm\sqrt{\l[(q-1)-c^2\rho^2\r]^2+2\kappa c^2\rho^2}\ ,\label{eq:vec_spectrum}\\
  \fl \frac{m_{\pm}^2}{H^2}&=\frac{\kappa}{q-1}\l[\frac{c^2}{H^2}+(p-1)\r]-(p-1)\pm\sqrt{(p-1)^2+2\kappa\frac{p-1}{q-1}\frac{c^2}{H^2}+\frac{2\kappa}{q-1}\frac{c^4}{H^4}}\ ,\nonumber
\end{eqnarray}
Like in the case of scalar perturbations, the set of eigenstates
$X^\pm_\mu$ that corresponds to the eigenvalues given in
equation~(\ref{eq:vec_spectrum}) consists of a linear combination of
the metric perturbations $V_{\mu n}$ and the flux perturbation
$b_{\mu n}$:
\begin{equation}
  \label{eq:mass_d}
  X^\pm_\mu=V_\mu+\frac{\kappa-m^2_\mp\rho^2 }{c^2\rho^2} cb_\mu\
  ,\spc\spc \Max X^\pm_\mu=m^2_\pm X^\pm_\nu\ ,
\end{equation}
where the equations are to be understood for each eigenfunction
$Y^I_{n}(y)$ separately with corresponding values of $\kappa^I$ and
$m^I_\pm$.

It can be shown~\cite{DeWolfe:2001nz} that the values of $\kappa$ are
bound from below by $\kappa\ge 2(q-1)$. The bound is saturated when
$Y_{(m|n)}=0$---i.e., for the Killing vectors of the compact space.
Together with this bound, it follows immediately from
equation~(\ref{eq:vec_spectrum}) that all the mass squares in the
spectrum are non-negative. Therefore, de Sitter compactifications are
stable with respect to vector modes, or in physical terms, the de
Sitter space is stable against the development of anisotropies.

The spectrum consists of two Kaluza--Klein towers, whose scaling
depends on two quantities: the size $\rho$ of the extra dimension
and the field strength of the flux $c$. It is of interest to
investigate the nature of the states with the lowest eigenvalues and
the spectrum in the limit of vanishing flux.

In the limit when the form flux is switched off, $c\rightarrow 0$,
its perturbation $b_\mu$ also vanishes. Only for the negative branch
do the coefficients of the mass eigenstates in
equation~(\ref{eq:mass_d}) remain regular and the limit can be
performed: $X_{\mu n}^-\propto V_{\mu n}$. The
spectrum~(\ref{eq:vec_spectrum}) reduces to
\begin{equation}
  \label{eq:vec_spectrum2}
  \rho^2m^2_-=\kappa-2(q-1)=\kappa-2(p-1)H^2\rho^2\ .
\end{equation}
The positive branch originates from the presence of the form field
fluctuations. It introduces a new scale $c$, but it does not affect
the lowest mass modes, which are in either case obtained from the
negative branch. One obtains $m_-^2=0$ for $\kappa=2(q-1)$ and all
values of $c$. The lowest vector modes are massless and correspond to
the isometries of the compact spacetime. The spectrum of vector modes
is summarized in Fig.~\ref{fig:vector_spectrum}.


In the absence of flux all remaining three parameters of the
background model $H$, $\Lambda$ and $\rho$ are related to each other
by factors of order unity, cf.\ equations~(\ref{eq:param_back}). The
flux introduces a new scale and enables us to create a hierarchy
between $H$ and the other parameters. When $c$ is very close to its
maximal value $c_\text{max}$, the number $H\rho^{-1}$ is very small.
This creates a hierarchy in a direction that is phenomenologically
interesting with $H\ll \rho^{-1}$. 
\begin{figure*}[h]
\begin{minipage}[t]{0.48\textwidth}
\rule{0.1em}{0em}\\
\hvFloat[nonFloat=true,capWidth=w]{figure}{
    \includegraphics[width=\columnwidth]{\Path{vector_spectrum_rho}}}{The
    dependence of the mass spectrum for the vector perturbations on
    the strength of the flux $c$ for $p=4$ and the compact space
    $S^3$.}{fig:vector_spectrum}
\end{minipage}\hfill
\begin{minipage}[t]{0.48\textwidth}
\rule{0.1em}{0em}\\
  \hvFloat[nonFloat=true,capWidth=w]{figure}{\includegraphics[width=\columnwidth]{\Path{tensor_spectrum_rho}}}{Spectrum of tensor fluctuations for $p=4$ and $S^3$.}{fig:tensor}
\end{minipage}
\end{figure*}
\label{fig:vector_tensor}

\subsection{Tensor Perturbations}
\label{sec:tensors}

\subsubsection{The Homogeneous Graviton.}

We show in this subsection that the homogeneous, $y$-independent
component of the graviton $h_{(\mu\nu)}$ reduces to the ordinary
$p$-dimensional de Sitter graviton. The details about the wave functions
and representations can be found in~\cite{Higuchi:1991tn}.

For the homogeneous mode all $y$-derivatives vanish. Then for the
Weyl-shifted metric perturbations $\delta
g_{\mu\nu}=h_{\mu\nu}-\frac{2q}{p-2}\Phi\gamma_{\mu\nu}$ and $\delta
g_{m n}=(1+2\Phi) g_{mn}$ the left-hand side of the Einstein
equations is obtained from equations~(\ref{eq:dricci})
\begin{equation}
  \label{eq:graviton_lhs_hom}
  \delta
  R^\text{hom}_{\mu\nu}=\frac{1}{2}\!\l(h_{\lambda\mu;\nu}^{\spc\spc\spc;\lambda}\!+\!h_{\lambda\nu;\mu}^{\spc\spc\spc;\lambda}\!-\!\Box h_{\mu\nu}\!-\!h^\lambda_{\lambda;\mu\nu}\r)\!+\!\frac{q}{p-2}\Box \Phi \gamma_{\mu\nu}\ .
\end{equation}
Similarly, the homogeneous contribution of the perturbations to the
$(\mu\nu)$-components of the energy--momentum tensor follows from
equation~(\ref{eq:ein4}):
\begin{equation}
  \label{eq:graviton_rhs_hom}
  \delta S^\text{hom}_{\mu\nu}=(p-1)H^2
  h_{\mu\nu}+2q\l(\frac{q-1}{p+q-2}c^2-\frac{p-1}{p-2}H^2\r)\Phi\gamma_{\mu\nu}\ .
\end{equation}
The terms in $\Phi$ of expressions~(\ref{eq:graviton_lhs_hom})
and~(\ref{eq:graviton_rhs_hom}) cancel by virtue of the equations of
motion~(\ref{eq:zeromode}) for the zero-mode of $\Phi$. Commuting
the derivatives in the expression~(\ref{eq:graviton_lhs_hom}) and
imposing transverse and traceless conditions on $h_{\mu\nu}$,
equations~(\ref{eq:graviton_lhs_hom})
and~(\ref{eq:graviton_rhs_hom}) reduce to the ordinary equation of
motion for the massless graviton in de Sitter space with an apparent
mass of $2H^2$:
\begin{equation}
  \label{eq:graviton_hom}
  \Box h_{\mu\nu}^\text{TT}=2H^2 h_{\mu\nu}^\text{TT}\ .
\end{equation}

\subsubsection{The Inhomogeneous Graviton.}

Slightly more involved is the identification of the massive
gravitons. The starting point is the traceless component of the
$(\mu\nu)$ Einstein equations:
\begin{eqnarray}
  \label{eq:tracefree}
\fl 0=
 \frac{1}{2}\l[h_{(\mu\lambda)\spc;\nu}^{\spc\spc\spc;\lambda}+ h_{(\nu\lambda)\spc;\mu}^{\spc\spc\spc;\lambda}-(\Box+\Delta) h_{(\mu\nu)}\r]\!+H^2 h_{(\mu\nu)}-\frac{1}{p} h_{(\kappa\lambda)}^{\spc\spc\spc;\kappa\lambda}\gamma_{\mu\nu}-(p-2)\Psi_{;(\mu\nu)}\ .
\end{eqnarray}
Next $h_{(\mu\lambda)}^{\spc\spc\spc;\lambda}$,
$h_{(\kappa\lambda)}^{\spc\spc\spc;\kappa\lambda}$ and $\Psi$ are
eliminated with the scalar
equations~(\ref{eq:scalar1})--(\ref{eq:scalar5}). One finds
\begin{equation}
  \label{eq:reduced_tracefree}
  \frac{1}{2}\l(\Box+\Delta\r)h_{(\mu\nu)}-H^2
  h_{(\mu\nu)}-2cb_{;(\mu\nu)}=0\ .
\end{equation}
The physical graviton is transverse and traceless. Following the
ansatz of~\cite{Kim:1985ez}, we construct the physical graviton
$\phi_{(\mu\nu)}$ as follows:
\begin{equation}
  \label{eq:phys_graviton}
  \phi_{(\mu\nu)}=h_{(\mu\nu)}-(ucb+v\Phi)_{;(\mu\nu)}\ ,
\end{equation}
where $u$ and $v$ are arbitrary constants, that are determined from
equation~(\ref{eq:reduced_tracefree}) and the conditions
\begin{eqnarray}
  \label{eq:transverse}
  \Box \phi_{(\mu\nu)}&=(-\Delta +2 H^2)\phi_{(\mu\nu)}\ ,\nonumber\\
  \phi^{(\mu\lambda)}_{\spc\spc\spc;\lambda}&=0 \ .
\end{eqnarray}
The value of the constants $u$ and $v$ depends on the eigenvalue of the
Laplacian $\Delta$ acting on the scalar functions $Y(y)$. They are
given by
\begin{eqnarray}
  \label{eq:coefficients}
   u=&\frac{2(p-2)}{(p-1)\l[(p-2)H^2-\lambda
     \rho^{-2}\r]}\ ,\nonumber\\
   v=&\frac{2(D-2)}{(p-1)\l[\lambda \rho^{-2}-(p-2)H^2\r]}\ ,
\end{eqnarray}
with the eigenvalues of the Laplacian $\lambda$ that can take values
$l(l+q-1)$ for $l\ge 0$.  From equation~(\ref{eq:transverse}) the
spectrum of the graviton is obtained. Neglecting the apparent mass
shift of $2H^2$ from the de Sitter space, one obtains a simple
Kaluza--Klein spectrum for the graviton modes, determined by the
geometry of the internal space and unaffected by the bulk matter
fields (in our case the form flux $F_{M_1\cdots M_q}$):
\begin{equation}
  \label{eq:mass_graviton}
  \rho^2 m^2_\text{grav} = l(l+q-1)\ ,\spc\spc\spc l\ge0\ .
\end{equation}

For comparison, the lowest Kaluza--Klein excitations of the
gravitational waves are plotted in Fig.~\ref{fig:tensor}.

\section{Physical Implications}
\label{sec:applications}

In this section we analyze two possible consequences for the
effective four-dimensional cosmology that result from the properties
of the mass spectrum of the scalar, vector and tensor perturbations
that we calculated in Sec.~\ref{sec:perturbations}.  (i) During
inflation an almost scale invariant spectrum of perturbations is
generated for modes whose mass is smaller than the scale of
inflation $H$. These perturbations contribute to the later evolution
of the universe. (ii) The perturbations gravitationally couple to
the standard model fields. We assume that the standard model fields
are realized as zero-modes of the corresponding higher dimensional
fields and calculate the coupling of the scalar perturbation to the
standard model fields.

\subsection{The Generation of Perturbations during Inflation}
\label{sec:inflation}

\begin{figure}[h]
  \centering
  \includegraphics[width=\textwidth]{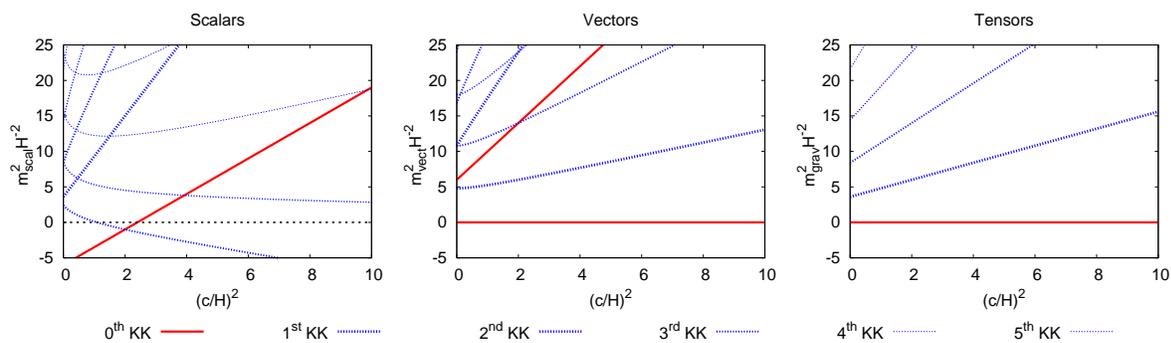}
  \caption{The masses of the scalar, vector and tensor
    perturbations in units of the expansion rate $H$ of the
    inflationary geometry for $p=4$ and $q=6$. Unlike the case in
    Figs.~\ref{fig:scalar_spectrum}--\ref{fig:vector_tensor},
    there is no limiting flux $c_\text{max}$ in this units. The flux
    $c$ can be arbitrarily larger than the expansions rate $H$.}
  \label{fig:masses_H}
\end{figure}

To estimate the dynamics of the perturbations during inflation it is
important to know their masses in comparison to the expansion rate
$H$. The perturbations are phenomenologically interesting when their
mass is smaller than the $H$. Figure~\ref{fig:masses_H} shows the
mass spectrum of the scalar, vector and tensor perturbations in
units of the expansion rate $H$ for a ten-dimensional spacetime
compactified on a six-dimensional sphere.

From the spectrum for the scalar perturbations in
Fig.~\ref{fig:masses_H} we see immediately that, besides the volume
modulus, higher Kaluza--Klein excitations can also have masses
smaller than the expansion rate $H$ for certain ranges of the form
field strength $c$. During inflation an almost scale invariant
spectrum of perturbations will be generated for these modes. They
therefore contribute as cosmological perturbations to the evolution
of the universe. In particular, we note the different nature of the
volume modulus and the higher Kaluza--Klein excitations with respect
to their coupling to the standard model fields, which is discussed
below, cf.~Sec.~\ref{sec:coupling}.

Apart from the massless vector fields that are associated with the
Killing vector fields on the $q$-sphere, all vector modes have
masses larger than the inflationary scale $H$. Consequently, the
Kaluza--Klein modes of vector perturbations are not excited during
inflation and should not play an important role in the subsequent
cosmological evolution.  The massless vector fields are conformal
and are not excited during inflation either. Moreover, they will
disappear from the spectrum, once compactifications to more
realistic internal spaces such as Calabi--Yau manifolds are
considered.

Similarly, the Kaluza--Klein excitations of tensor modes have masses
above the scale of inflation and do not contribute to the dynamics
of cosmological perturbation after inflation. The massless tensor
mode corresponds to the ordinary four-dimensional graviton. It
carries no information about the extra dimensional nature of the
full spacetime. Similar conclusions were derived for braneworld
models in~\cite{Frolov:2002qm}.

\subsection{The Coupling to Standard Model Fields}
\label{sec:coupling}

Next, we briefly discuss the nature of the coupling of the scalar
perturbations to the standard model fields. As a consequence of
their different nature, the zero-modes and higher Kaluza--Klein
excitations couple differently to the standard model fields. When
the standard model fields are localized on a brane, it is known that
the radion universally couples to the trace of the energy--momentum
tensor of the brane degrees of freedom:
\begin{equation}
  \label{eq:brane_coupling}
  S^\text{brane}_\text{int}=\int d^4x \sqrt{\gamma} \Phi T^\mu_{\,\mu}\ .
\end{equation}
However, in Kaluza--Klein compactifications one does not need to
assume that the standard model fields are localized. Instead, the
standard model fields of the effective four-dimensional theory
simply correspond to the zero-modes of the higher dimensional
fields.  This point of view modifies the coupling of the volume
modulus to the standard model zero-modes and
equation~(\ref{eq:brane_coupling}) is not applicable.

We start with the action~(\ref{eq:action}) and add the standard
model Lagrangian as a collection of zero-modes to the full higher
dimensional theory:
\begin{equation}
  \label{eq:SM_setup}
  S=\int d^4x d^q y\sqrt{g_{4+q}}\l\{\frac{1}{2}R -
    \frac{1}{2q!} F^2_{q}-\Lambda+{\cal L}^\text{zero}_\text{SM}\r\}\
    .
\end{equation}
For simplicity, we only analyze a canonical scalar field $\chi(x)$:
\begin{equation}
  \label{eq:SM_scalar}
  {\cal L}^\text{zero}_\text{SM}=-\frac{1}{2}\l(\partial_\mu \chi\r)^2-\frac{1}{2}m^2\chi^2\ .
\end{equation}
Focusing on the scalar perturbations $\hat \Psi$ and $\Phi$
introduced in the line element~(\ref{eq:line}), the
action~(\ref{eq:SM_setup}) is expanded to first order:
\begin{eqnarray}
  \label{eq:SM_linear}
 \fl S=\int d^4x
  d^qy\sqrt{\gamma}\sqrt{g}\l(1\!+\!4\hat\Psi\!+\!q\Phi\r)\l\{\frac{1}{2}(1\!-\!2\hat \Psi)R[\gamma]+\cdots\r.\nonumber\\\l.-\frac{1}{2}\l(1\!-\!2\hat \Psi\r)(\partial_\mu \chi)^2-\frac{1}{2}m^2\chi^2\r\} \ ,
\end{eqnarray}
where the dots collect the terms from the form fields, the
cosmological constant and the curvature of the internal space. Next
the Weyl shift $\hat \Psi=\Psi-\frac{q}{2}\Phi$, cf.\
equation~(\ref{eq:weylshift}), is performed to obtain the linearized
Einstein frame and the extra dimensions are integrated out:
\begin{eqnarray}
  \label{eq:effective4d_general}
  \fl S_\text{eff}=\int d^4x
  \sqrt{\gamma}&\l\{\frac{1}{2}\l[1+2\Psi\r]\l[\frac{1}{M_\text{P}^2}R-\l(\partial_\mu\chi\r)^2\r]+\cdots-\frac{1}{2}\l(1+4\Psi-q\Phi\r)m^2\chi^2\r\}\ ,
\end{eqnarray}
where the four-dimensional Planck mass $M_\text{P}$ arises from the
rescaling of the fundamental scale with the volume of the internal
space.  Similarly the field $\chi$ is rescaled by the volume of the
internal space $\sqrt{V_q}\chi\rightarrow\chi$. The term
$\frac{qm^2}{2}\Phi\chi^2$ is the reason that the coupling of the
volume modulus to the standard model fields deviates from the form
in equation~(\ref{eq:brane_coupling}). Apart from this term the
action~(\ref{eq:effective4d_general}) corresponds to a
four-dimensional theory of gravity and a canonical scalar field
$\chi$ in a de Sitter geometry with metric fluctuations of the form
$ds^2=(1+2\Psi)\gamma_{\mu\nu}dx^\mu dx^\nu$.

We now discuss how the scalar perturbations interact with the
standard model fields for the two different cases of a homogeneous
(volume modulus) and inhomogeneous Kaluza--Klein excitations.

\subsubsection{The Coupling of the Homogeneous Volume Modulus.}

The result of the zero-mode for scalar perturbations derived in
Sec.~\ref{sec:scalars} was particularly simple:
\begin{eqnarray}
  \label{eq:zeromode2}
  \Psi&=&0\ ,\nonumber\\
  m^2_{\Phi}&=& -6H^2+\frac{4(q-1)}{q+2}c^2\ .
\end{eqnarray}
Therefore, we obtain the effective four-dimensional action for the
zero-mode:
\begin{eqnarray}
  \label{eq:effective4d}
  \fl S_\text{eff}=\int d^4x
  \sqrt{\gamma}&\l\{
  \frac{M_\text{P}^2}{2} R-\frac{1}{2}\l[\l(\partial_\mu\chi\r)^2+{\cal
    N}(\partial_\mu \Phi)^2\r]-\frac{1}{2} (m^2\chi^2+{\cal N}m^2_\Phi \Phi^2)+\frac{qm^2}{2M_\text{P}}\Phi\chi^2\r\}\ ,\nonumber\\
\end{eqnarray}
where $\Phi$ is rescaled to its canonical form up to a constant ${\cal
  N}$ of order unity. The effective action shows the volume modulus as a
canonical scalar field with mass $m_\Phi$ that couples to the
standard model scalar fields through the interaction term
$\Phi\chi^2$ and the Planck suppressed coupling $qm^2/M_\text{P}$.
In particular, it has no first-order coupling at all to massless
(conformal) fields.

\subsubsection{Non-zero Mode Coupling.}

Now one can use equation~(\ref{eq:scalar3}) to eliminate $\Psi$. The
following four-dimensional interactions are obtained from the
effective action~(\ref{eq:effective4d_general}):
\begin{eqnarray}
  \label{eq:interactions}
  S_\text{int}\propto\int\!
  d^4x\sqrt{\gamma}&\l\{\frac{q-2}{4}\frac{\Phi}{M_\text{P}}\l[M^2_\text{P}R-(\partial_\mu \chi)^2\r]\!+\!\frac{\Phi}{M_\text{P}}m^2\chi^2\r\}\ .
\end{eqnarray}
However, one has to take into account that $\Phi$ is not the
physical dynamical variable for the massive modes. The correct
kinetic terms and normalizations have to be found for the mass
eigenstates calculated in equation~(\ref{eq:scalar_mass}), which are
linear combinations of the scalar perturbation $\Phi$ and the scalar
matter degrees of freedom, in our case the scalar component of the
form flux $b$.  Qualitatively, the interactions in
equation~(\ref{eq:interactions}) show the direct but Planck mass
suppressed decay channel of massive Kaluza--Klein states into
gravitons and standard model fields.  Alternatively, Kaluza--Klein
excitations of the scalar perturbation $\Phi$ around its expectation
value lead to variations of the Planck mass and the masses of the
standard model fields.

\section{Conclusions}

In this paper, we systematically analyzed the stability properties
of de Sitter compactifications with $q$-form fluxes. We calculated
in a unified way the complete perturbative mass spectrum of de
Sitter compactifications. The most important feature of the
perturbative spectrum is the appearance of tachyonic modes in the
spectrum of scalar perturbations.  The remaining masses of the
vector and tensor perturbations of the spectrum are non-negative and
therefore do not create additional instabilities.

The tachyonic mode of the volume modulus, possible ways of
stabilizing it, and implications for inflation have been discussed
before in~\cite{Frolov:2003yi,Contaldi:2004hr}. Its occurrence
imposes tight constraints on the maximal scale of inflation. It also
implies that the volume of the internal space is stabilized at least
at the scale of inflation to ensure a stable background
configuration and a sufficiently large number of efolds during
inflation. If the stabilization of the volume modulus remains at
such a high scale after inflation, the modulus is too heavy to be
detected in future accelerator experiments.  The tachyonic nature of
the unstabilized modulus is not related to special properties of the
internal space. It merely reflects the presence of the inflationary
geometry with constant expansion rate $H$ in the effective
four-dimensional theory.

Further tachyonic modes that can arise from the quadrupole and
higher Kaluza--Klein excitations are less explored. Non-perturbative
dynamics that are triggered by these instabilities are not known. A
possible stabilization due to additional matter fields is far from
obvious, since the instability originates from the presence of the
form flux. We therefore expect similar obstructions caused by this
instability in more general compactifications.

\begin{figure}[h]
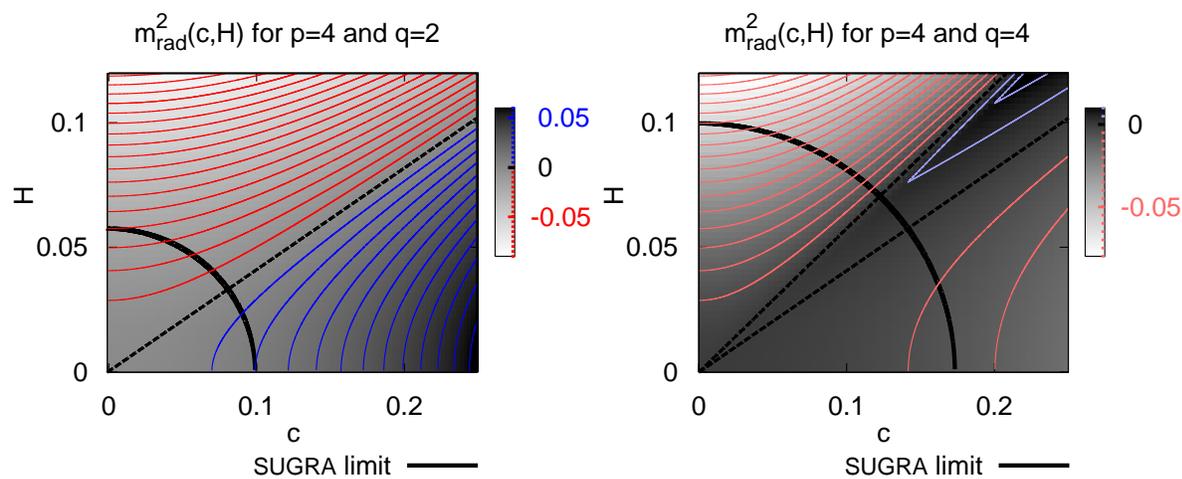

\begin{minipage}{\textwidth}
\begin{minipage}[t]{0.5\textwidth}
\begin{flushleft}
  \includegraphics[width=\columnwidth]{\Path{radion_mass2}}\\
\end{flushleft}
\end{minipage}\hfill
\begin{minipage}[t]{0.5\textwidth}
\begin{flushright}
   \includegraphics[width=\columnwidth]{\Path{radion_mass1}}\\
\end{flushright}
\end{minipage}
\end{minipage}
\caption{The mass of the lowest lying scalar excitation is shown for two
  extra dimensions in the left panel and for four extra dimensions in
  the right panel. The contour lines have a separation of $0.005$ (in
  fundamental units). The blue (red) contour lines indicate regions of
  (un)stable compactifications.  In the left panel the dashed line
  represents the parameters for which the mass of the volume modulus
  vanishes. In the right panel the upper dashed line indicates the
  zeros for the mass of the volume modulus and the lower dashed line
  shows where the mass of the second KK mode vanishes.  Stable
  compactifications are only possible for parameters in between the two
  dashed lines.  The solid black line encircles the region in which the
  size of the extra dimensions is larger than $10$ fundamental units
  of lengths (i.e.,\ the valid region for the supergravity
  approximation).}
\label{fig:radion_mass}
\end{figure}

Figure~\ref{fig:radion_mass} shows the contour plots for the value
of the smallest mass of scalar perturbations as a function of the
Hubble scale $H$ and form flux $c$. All quantities are plotted in
units of the fundamental mass scale $M=1$.  For two or three extra
dimensions the lowest mass is by and large determined by the volume
modulus, cf.\ equation~(\ref{eq:radion}). For four and more extra
dimensions the negative branch in equation~(\ref{eq:scalar_mass}) of
the higher Kaluza--Klein excitations becomes tachyonic for a
sufficiently large contribution from the form flux. For four extra
dimensions only a small range of parameters $H$ and $c$ admits
stable compactifications.  For more than four extra dimensions no
stable compactifications are found at all in this model. The solid
black line in Fig.~\ref{fig:radion_mass} encircles the region where
the size of the extra dimensions $\rho > 10M^{-1}$. It is called the
SUGRA limit, since for smaller values of $\rho$---i.e., for
parameters $H$ and $c$ outside the encircled region---quantum
gravity corrections are expected to contribute.

We investigated the calculated spectrum of scalar, vector and tensor
perturbations for possible phenomenological consequences. For
certain ranges of the form field strength $c$, the scalar sector
provides light modes that generate an almost scale-invariant
spectrum of fluctuations during inflation. These perturbations decay
gravitationally into standard model fields after inflation. The
details of the decay are different for the volume modulus and the
higher Kaluza--Klein excitations.

The spectrum of vector perturbations~(\ref{eq:vec_spectrum}) does
not contain tachyonic modes. Moreover, apart form the massless gauge
fields that correspond to the Killing vector fields on the
$q$-sphere, all vectors have a mass larger than the scale of
inflation, $m_\text{vec}>H$ .  Furthermore, the vector modes do not
have a direct decay channel into standard model fields, since the
only conceivable first-order coupling to the standard model fields
of the form $V_{(\mu;\nu)} T_{SM}^{\mu\nu}$ can be integrated by
parts and vanishes due to the properties of the energy--momentum
tensor.

As expected, the spectrum of tensor perturbations does not depend on
the details of the matter field content. It only measures the
geometry of the compactification. The zero-mode does not feel the
presence of the extra dimensions and behaves like an ordinary
graviton in the de Sitter geometry.  The Kaluza--Klein excitations
of the graviton only depend on the size and the shape of the extra
dimensions. Again, the massive modes have masses larger than the
scale of inflation, $m_\text{tensor}>H$.  Therefore, Kaluza--Klein
excitations of tensor modes will not be generated during inflation.
Possible existing excitations from physics before inflation are
erased during the process of inflation.  Consequently, the massive
gravitons are not expected to play a role for cosmological
observations.

In the final remark we want to compare this spectrum to the spectrum
of anti-de Sitter compactifications calculated in~\cite{Kim:1985ez}.
The main difference in the set up between the two compactifications
is the sign in the curvature of the large dimensions and the
additional bulk cosmological constant. However, many features do not
change qualitatively.  Most importantly, the mixing between matter
and higher dimensional metric degrees of freedom in the scalar and
vector sector is common to both compactifications. It is a well
known phenomenon with importance also for the ordinary theory of
inflation~\cite{Mukhanov:1992me} and braneworld scenarios. The
mixing between the higher dimensional metric perturbations and
matter fields in braneworlds was investigated in detail
in~\cite{Kofman:2004tk}. Even the appearance of modes with negative
mass square is common to both supersymmetric and non-supersymmetric
compactifications. However, in the case of anti-de Sitter
compactifications the stability against perturbations with tachyonic
mass is ensured by the Breitenlohner-Freedman
bound~\cite{Breitenlohner:1982jf}, as long as their mass is larger
than this bound.

\section*{Acknowledgement}

This paper is part of my Ph.D. thesis about ``Cosmology with Extra
Dimensions''. I am particularly grateful to my supervisor Lev Kofman
for guidance. Additionally, I thank Marco Peloso, Erich Poppitz and
Pascal Vaudrevange for useful discussions. Support from the Connaught
Scholarship is acknowledged.

\appendix

\section{Notation}
\label{sec:Notation}
\begin{tabular}[t]{l l}
  $x^M$ &: $(p+q)$-dimensional set of coordinates\\
  $x^\mu$ &: de Sitter space coordinates\\
  $y^m$ &: Coordinates of the compact space\\
  $G_{MN}$ &: Metric of the full spacetime\\
  $\gamma_{\mu\nu}(x)$&: de Sitter Metric\\
  $g_{mn}(y)$ &: Metric of the compact space\\
  $\nabla_M\phi$ &: Covariant derivative that preserves $G_{M\!N}$\\
  $\nabla_\mu\phi\!\equiv\!\phi_{;\mu}$ &: Covariant derivative that preserves
  $\gamma_{\mu\nu}$\\
  $\nabla_n \!\phi\!\equiv\!\phi_{|n}$ &: Covariant derivative that preserves
  $g_{mn}$\\
  $Y^{k}(y)$ &: Scalar harmonics of the compact space\\
  $Y^{k}_{m}(y)$ &: Vector harmonics of the compact space\\
  $Y^{k}_{(mn)}(y)$ &: Transverse and traceless harmonics\\
\end{tabular}

\section{Weyl shift}
\label{sec:weyl}

To understand the redefinition of the scalar perturbation $\hat \Psi$
in equation~(\ref{eq:weylshift}), we consider the effective
$p$-dimensional theory that is obtained from the compactification of a
$(p+q)$-dimensional theory with a choice of metric of the form
\begin{equation}
  \label{eq:weyl_metric}
  g_{MN}=\l(
  \begin{array}{c c}
    \gamma_{\mu\nu} & 0 \\
    0 & e^{2\Phi} g_{mn}
  \end{array}\r ) \ .
\end{equation}
For the zero-mode of $\Phi$ and this choice of metric the $(p+q)$-dimensional
action of gravity reduces to
\begin{equation}
  \label{eq:weyl_einstein}
  \int\!d^px d^qy\sqrt{|g_{p+q}|} R_{p+q}\!=\!V_q\! \int\! d^px \sqrt{|
    \gamma|}e^{q\Phi}\l[R_{\gamma}\!+\!e^{-2\Phi} R_g\r].
\end{equation}
To obtain a canonically normalized four dimensional graviton or, in
other words, a four-dimensional Einstein theory of gravity, the
metric has to be rescaled by a Weyl transformation:
\begin{equation}
  \label{eq:weyl_trafo}
  \gamma_{\mu\nu}\rightarrow \exp\l({\frac{2q}{p-2}\Phi}\r)\gamma_{\mu\nu}\ .
\end{equation}
If the field $\Phi$ is treated as a perturbation as in
equation~(\ref{eq:line}), the above Weyl
transformation~(\ref{eq:weyl_trafo}) of the metric
$\gamma_{\mu\nu}$ amounts to the Weyl shift~(\ref{eq:weylshift}) of
the field $\hat\Psi$.

\section{Form Equations}
\label{sec:formequations}

In this appendix we list useful formulae that have been used to
simplify the equations of motion. They are entirely based on the
properties of the antisymmetric epsilon tensor and are
straightforward to derive:
\begin{eqnarray}
  \label{eq:formulae}
  f_{m_1\cdots m_q}\epsilon^{m_1\cdots m_q}&=&q!\, \Delta b\ ,
  \nonumber\\
  f_{m m_2\cdots m_q}{\epsilon_n}^{m_2\cdots
    m_q}&=&(q\!-\!1)!\,\Delta b\,g_{mn}\ ,\\
  f_{\mu m_2\cdots m_q}\epsilon_n^{\,\,m_2\cdots
    m_q}&=&(q\!-\!1)!\l(b_{;\mu|n}+{\hat\Delta^l}_{\,\,n} b_{l\mu}\r) .\nonumber
\end{eqnarray}

\section{Residual Gauge Freedom}
\label{sec:residual}

As mentioned in Sec.~\ref{sec:class_gauge}, the de Donder gauge
conditions~(\ref{eq:dedonder}) do not fix the gauge freedom
associated with the infinitesimal coordinate
transformations~(\ref{eq:inf_coord}) completely. Consequently,
there will be modes in the spectrum of perturbations that do not
correspond to physical degrees of freedom. In this section, we analyze
the nature of these residual gauge degrees and impose additional
constraints to eliminate them.

The residual gauge freedom consists of functions that satisfy the
additional constraints
\begin{eqnarray}
  \label{eq:app_residual}
  \Delta \xi_\mu+{\xi^l}_{|l;\mu}&=&0\ ,\nonumber\\
  \xi_{(m|n)}&=&0\ .
\end{eqnarray}
There are three distinct solutions that satisfy these constraints:

\paragraph{The $y$-independent infinitesimal diffeomorphisms}
\begin{eqnarray}
  \label{eq:hom_diff}
  \xi^\mu(x)&=\xi^\mu_\perp(x)+\xi(x)^{;\mu}\nonumber\ ,\\
  \xi^l(x)& \ ,
\end{eqnarray}
where we split the $y$-independent diffeomorphisms $\xi^\mu(x)$ into
a transverse vector $\xi^\mu_\perp$ and a scalar function $\xi(x)$.
Similarly, we decompose the traceless part of homogeneous metric
perturbations:
\begin{equation}
  \label{eq:hom_metric}
  h_{(\mu\nu)}(x)=h^\text{TT}_{(\mu\nu)}(x)+F^\perp_{(\mu;\nu)}(x)+E_{;(\mu\nu)}(x)\ ,
\end{equation}
into the transverse-traceless polarizations
$h^\text{TT}_{(\mu\nu)}$, the transverse vector $F^\perp_{\mu}$ and
the scalar function $E$. Correspondingly, the homogeneous
off-diagonal components of the metric perturbations are decomposed:
\begin{equation}
  \label{eq:{eq:hom_off}}
  V_{\mu n}(x)=V^\perp_{\mu n}+B_{n;\mu}\ ,
\end{equation}
into transverse and longitudinal polarizations.

From the standard transformation of the homogeneous metric
perturbations under infinitesimal transformations
\begin{equation}
  \label{eq:trans}
  \delta g_{MN}(x)\rightarrow \delta g_{MN}(x) -\pounds_\xi
  g_{MN}(x)\ ,
\end{equation}
with $\pounds_\xi$ denoting the Lie derivative in direction of
$\xi^M$, follows that the additional gauge constraints can be
imposed:
\begin{eqnarray}
  \label{eq:res_constraint}
  F^\perp_{\mu}=0\ ,\spc\spc\spc E=0\ ,\spc\spc\spc B_n=0\ .
\end{eqnarray}
These additional constraints ensure that the homogeneous graviton
$h_{(\mu\nu)}$ only contains transverse-traceless polarizations and
that the homogeneous metric components $V_{\mu n}$ are transverse
vectors.

\paragraph{The Killing vectors of the $q$-sphere}
$[0,\xi^l_{(a)}(y)]$ are another set of solutions $\xi^M_{(a)}$,
$a=1,\cdots, \frac{1}{2}q(q+1)$ to the
equations~(\ref{eq:app_residual}). This gauge symmetry is not removed
by further constraints. It remains a symmetry of the effective
theory, provided that the massless vector fields $V_{\mu n}$ transform
in the adjoint representation of the isometry group of the $q$-sphere,
which is generated by the Killing vectors $\xi^l_{(a)}$.

\paragraph{The conformal diffeomorphisms} are generated by the
scalar harmonics on the $q$-sphere $Y^I(y)$ with $l=1$ (i.e., the
scalar harmonics that correspond to the first level of Kaluza--Klein
excitations). They satisfy the additional constraint
$Y_{|(mn)}\equiv Y_{|mn}-\frac{1}{q}g_{mn} \Delta Y=0$. The solution
to the equations~(\ref{eq:app_residual}) is given by
\begin{equation}
  \label{eq:conformal_diff}
  \xi_M=[-k^I_{;\mu}(x)Y^I(y),k^I(x) Y^I_{|m}(y)]\ .
\end{equation}
This residual symmetry is the reason why the negative branch of the
spectrum~(\ref{eq:scalar_mass}) does not contribute to the physical
degrees of freedom of the first excited Kaluza--Klein modes (i.e.,
$l=1$)~\cite{vanNieuwenhuizen:1984iz}.

\providecommand{\href}[2]{#2}\begingroup\raggedright

\endgroup

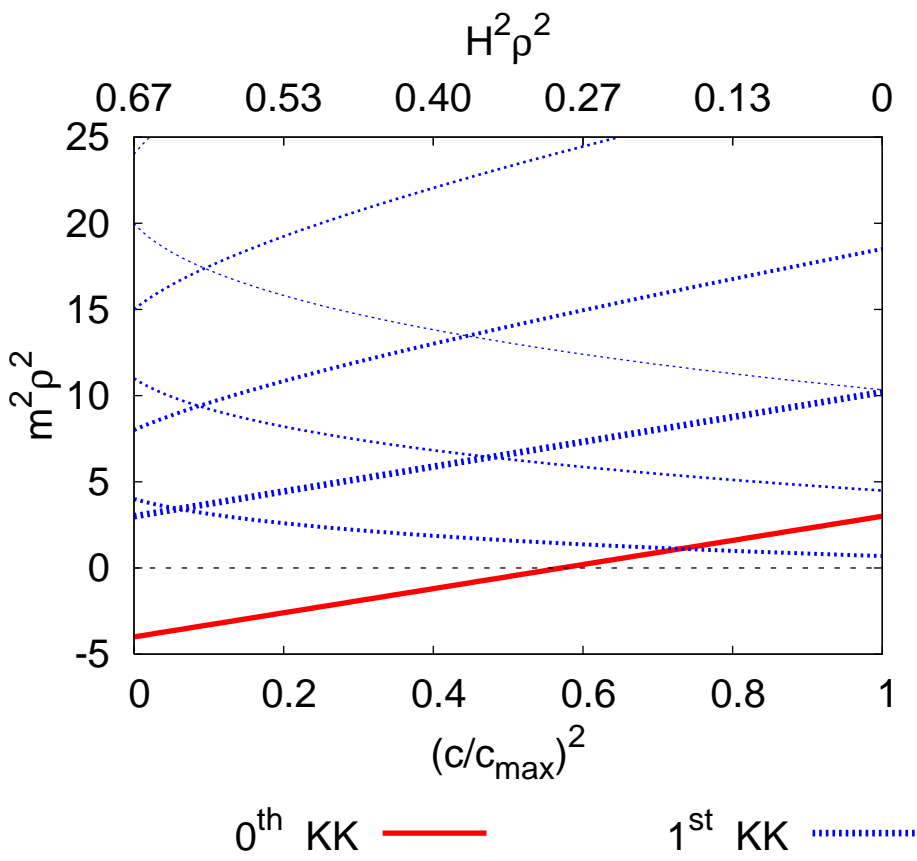
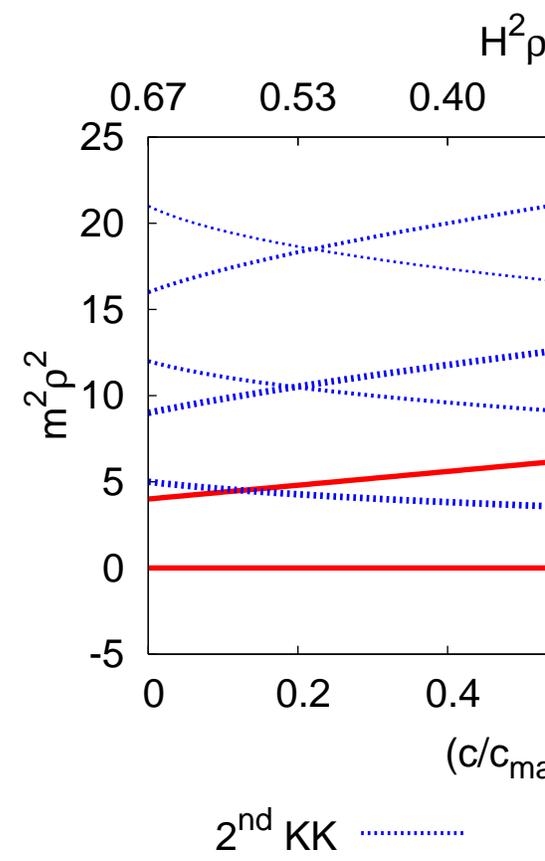
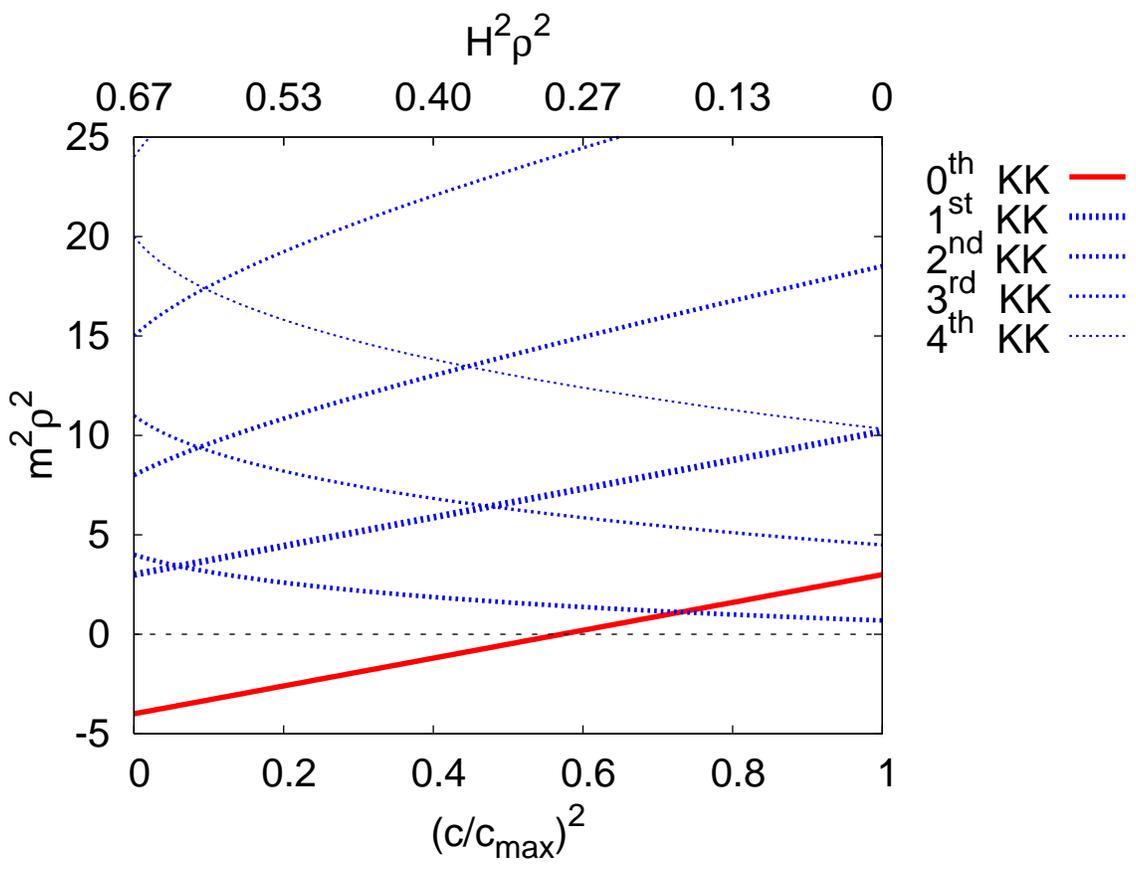
\end{document}